\documentclass[preprint,12pt]{elsarticle}

\usepackage{amsmath,amsfonts,amssymb,amsthm}
\usepackage{lipsum,appendix}

\newcommand{\nop}[1]{}
\newtheorem{theorem}{Theorem}
\newtheorem{lemma}{Lemma}
\newtheorem{definition}{Definition}
\newtheorem{corollary}{Corollary}

\newtheorem{example}{Example}

\journal{Performance Evaluation}

\begin{document}

\begin{frontmatter}



\title{A Temporal Approach to Stochastic Network Calculus \footnote{An early version of this paper was partially presented at MASCOTS 2009 \cite{Xie:timedomain}.}}

\author[label1]{Jing Xie}
\author[label2]{Yuming Jiang}
\author[label3]{Min Xie}
\address[label1]{Research and Innovation, Det Norske Veritas\\
Veritasveien 1, 1363, H\o vik, Norway}
 \address[label2]{Centre for Quantifiable Quality of Service in Communication Systems (Q2S) and \\
Department of Telematics\\
Norwegian University of Science and Technology (NTNU), Norway}
 \address[label3]{University Centre at Blackburn College\\
University Close, United Kingdom}



\begin{abstract}
Stochastic network calculus is a newly developed theory for stochastic service guarantee analysis of computer networks. In the current stochastic network calculus literature, its fundamental models are based on the cumulative amount of traffic or cumulative amount of service. However, there are network scenarios where direct application of such models is difficult. This paper presents a temporal approach to stochastic network calculus. The key idea is to develop models and derive results from the time perspective. Particularly, we define traffic models and service models based on the cumulative packet inter-arrival time and the cumulative packet service time, respectively. Relations among these models as well as with the existing models in the literature are established. In addition, we prove the basic properties of the proposed models, such as delay bound and backlog bound, output characterization, concatenation property and superposition property. These results form a temporal stochastic network calculus and compliment the existing results. 
\end{abstract}


\end{frontmatter}


\section{Introduction}\label{Sec-Intro}

Stochastic network calculus is a theory dealing with queueing systems found in computer networks \cite{Change:ExStoLinMax}\cite{Fidler:EndMGF}\cite{Jiang:BasicStoNet}\cite{Jiang:book}. It is particularly useful for analyzing networks where service guarantees are provided stochastically. Such networks include wireless networks, multi-access networks and multimedia networks where applications can tolerate a certain level of violation of the desired performance \cite{Ferrari:Service}.  

Stochastic network calculus is based on properly defined traffic models \cite{Change:traffic}\cite{Jiang:BasicStoNet}\cite{Jiang:book}\cite{Li:NC2EB}\cite{David:SBB}\cite{Yaron:traffic} and service models \cite{Jiang:BasicStoNet}\cite{Jiang:book}. In literature, it is typical to model the arrival process by a stochastic arrival curve and the service process by a stochastic service curve. The arrival curve provides probabilistic upper bounds on the cumulative amount of arrival traffic whereas the service curve lower bounds the cumulative amount of service. In this paper, we call such models {\em space-domain} models, from which extensive results have been derived. There are five most fundamental properties \cite{Jiang:BasicStoNet}\cite{Jiang:book}:~(P.1)~{\em Service Guarantees} including delay bound and backlog bound;~(P.2)~{\em Output Characterization};~(P.3)~{\em Concatenation Property}; (P.4) {\em Leftover Service}; (P.5) {\em Superposition Property}. Examples demonstrating the necessity and applications of these basic properties can be found \cite{Jiang:BasicStoNet}\cite{Jiang:book}.

However, there are many open challenges for stochastic network calculus, making its wide application difficult \cite{Jiang:book}. One is to analyze networks where users are served probabilistically. For example, in wireless networks, a wireless link is error-prone and consequently retransmission is often adopted to ensure reliability. In random multi-access networks, random backoff and retransmission are used to deal with contention and collision. To apply stochastic network calculus to analyze such networks, it is fundamental to find the stochastic characterization of the service time that provides successful transmissions for the user. However, direct application of existing space-domain models, which are built on the amount of cumulative service, is difficult. 

This paper aims to rethink stochastic network calculus to address some of the challenges in current stochastic network calculus literature, such as the analysis of error-prone wireless channels and/or contention-based multi-access. To be specific, we present a temporal approach to stochastic network calculus. The key idea is to develop models and derive results from the time perspective. We define traffic models and service models based on the cumulative packet inter-arrival time and the cumulative packet service time, respectively. In this paper, we shall call such models {\em time-domain} models. In addition to their easy use in network scenarios discussed above, the basic properties are also investigated based on the proposed time-domain models. Moreover, relations among the proposed {\em time-domain} models as well as with the corresponding {\em space-domain} models are established, which provide a tight link between the proposed temporal stochastic network calculus approach and the existing space domain stochastic network calculus approach. This gives increased flexibility in applying stochastic network calculus in challenging network scenarios.

The structure of the paper is as follows. In Section \ref{Sec-Background} we first introduce the notation and the system specification, followed by the review of the relevant results of stochastic network calculus. Section \ref{Sec-ModelDefinition} defines the network calculus models in the time-domain and explores the model transformations. Four fundamental properties\nop{, \emph{service guarantees, output characterization, concatenation property and superposition property},} are thoroughly investigated in Section \ref{Sec-Property}. The relevant discussion reveals the reasons of establishing the model transformations in Section \ref{Sec-ModelDefinition}. In Section \ref{Sec-Conclusion}, we conclude the paper and discuss the open issue.

\section{Network Model and Related Work}\label{Sec-Background}

This section specifies the network system and reviews mathematical preliminaries for the analysis in the following sections. A brief overview on stochastic network calculus of particular relevance to this paper is presented as well.

In this paper, we make the following assumptions unless stated otherwise. 
\begin{itemize}
  \item All packets have the same length.
  \item A packet is considered to be received by a network element when and only when its last bit has arrived to the network element. 
  \item A packet can be served only when its last bit has arrived. 
  \item A packet is considered out of a network element when and only when its last bit has been transmitted by the network element. 
  \item Packets arriving to a network element are queued in the buffer and served in the FIFO order. All queues are \nop{assumed to be }empty at time $0$.
  \item All network elements provide sufficient buffer space to store all incoming traffic and are lossless. 
\end{itemize} 
\nop{A process of characterizing the network temporal behavior is defined to be a function of packet sequence number n ($n\ge 0$). In this paper, we consider packets arriving to a network system according to some general inter-arrival time distribution. }

\subsection{Notations and System Specification}\label{Subsec-Notation}

We use $P(n)$, $a(n)$, $d(n)$ and $\delta_n$, to denote the $(n+1)th$ packet entering the system, its arrival time to the system, its departure time from the system and its service time provided by the system, respectively, where $n=0,1,2,...$. 


\begin{itemize}
  \item From the temporal perspective, an arrival process counts the cumulative inter-arrival time between two arbitrary packets and is denoted by $\Gamma(m,n)=a(n)-a(m)$ for any $0\le m\le n$. Note $\Gamma(n,n)=0$.
  \item A service process describes the cumulative service time received between two arbitrary packets and is denoted by $\Delta(m,n)=\sum_{k=m}^n\delta_k$ for any $0\le m\le n$. Note $\Delta(n,n)=\delta_n$.
\end{itemize}    

\nop{By convention, all processes are defined on $0\le m\le n$. In this paper, both $a(n)$ and $\Gamma(m,n)$ are used interchangeably to represent an arrival process. }


In the time-domain, the system backlog and system delay are defined below, respectively. 
\begin{definition}
\nop{Let $a(n)$ and $d(n)$ be the time of packet $P(n)$ arriving to a system and that of departing from the system, respectively. Let the departure time of packet $P(n)$ be $d(n)=t$ ($t\ge 0$). Then} The system backlog at time $t\ge 0$ is denoted by $B(t)$: 
\begin{eqnarray}
B(t)\le \inf\Big\{l\ge 0,\sup\{n\ge 0: a(n)\le t\}: d(n-l)\le a(n)\Big\}.\label{eq:timebacklog}
\end{eqnarray}
The delay that packet $P(n)$ experiences in the system is denoted by $D(n)$:
\begin{eqnarray}
D(n) = d(n) - a(n). \label{eq:timedelay}
\end{eqnarray} 
\end{definition}
 
Moreover, the time that packet $P(n)$ waits in queue is denoted by $W(n)$:
\begin{eqnarray}
W(n) = D(n) - \delta_n. \label{eq:queueingdelay}
\end{eqnarray} 

The following function sets are often used in this paper. 

The set of non-negative wide-sense increasing functions is denoted by $\mathcal{F}$, where for each function $f(\cdot)$, 
\begin{displaymath}
\mathcal{F}=\big\{f(\cdot): \forall 0\leq x\leq y, 0 \leq f(x) \leq f(y)\big\}
\end{displaymath}  
and for any function $f(\cdot)\in\mathcal{F}$, we set $f(x)=0$ for all $x<0$. 

We denote by $\bar{\mathcal{F}}$ the set of non-negative wide-sense decreasing functions where for each function $f(\cdot)$, 
\begin{displaymath}
\bar{\mathcal{F}}=\big\{f(\cdot): \forall 0\leq x\leq y, 0 \leq f(y) \leq f(x)\big\}
\end{displaymath}  
and for any function $f(\cdot)\in\bar{\mathcal{F}}$, we set $f(x)=1$ for all $x<0$. 

We denote by $\bar{\mathcal{G}}$ a subset of $\bar{\mathcal{F}}$, where for each function $f(\cdot)\in\bar{\mathcal{G}}$, its nth-fold integration, denoted by $f^{(n)}(x)\equiv(\int_x^\infty ~dy)^nf(y)$, is bounded for any $x\ge 0$ and still belongs to $\bar{\mathcal{G}}$ for any $n\ge 0$, i.e., 
\begin{eqnarray}
\bar{\mathcal{G}} = \big\{f(\cdot): \forall n\ge 0, (\int_x^\infty ~dy)^nf(y) \in \bar{\mathcal{G}}\big\}. \nonumber
\end{eqnarray}

For ease of exposition, we adopt the following notations in this paper: $$[x]^+\equiv \max[x,0] ~~\text{and}~~ [x]_1\equiv \min[x,1].$$

In addition, the ceiling and floor functions are used in this paper as well. 
\begin{itemize}
  \item The ceiling function $\lceil x\rceil$ returns the smallest integer not less than $x$. 
  \item The floor function $\lfloor x\rfloor$ returns the larget integer not greater than $x$.  
\end{itemize}

\subsection{Mathematical Basis}\label{Subsec-math}

An essential idea of (stochastic) network calculus is to use alternate algebras, particularly min-plus algebra and max-plus algebra \cite{Boudec:DeterNC}, to transform complex non-linear network systems into analytically tractable linear systems. To the best of our knowledge, the existing models and results of stochastic network calculus mainly based on min-plus algebra that has basis operations suitable for characterizing the amount of \emph{cumulative} traffic and service. As a result, these models focus on describing network behavior from the spatial perspective. Max-plus algebra is suitable for arithmetic operations with cumulative inter-arrival times and service times. Consequently, network modeling from the temporal perspective more relies on max-plus algebra. In the following, we review the basics of both min-plus algebra and max-plus algebra.

In min-plus algebra, the \lq addition\rq~operation represents \emph{infimum} or \emph{minimum} when it exists, and the \lq multiplication\rq~operation is $+$. The \emph{min-plus convolution} of functions $f, g\in\mathcal{F}$, denoted by $\otimes$, is defined as  
  \begin{displaymath}
     (f\otimes g)(t) = \inf_{0\le s\le t}\{f(s)+g(t-s)\}
  \end{displaymath}
where, when it applies, \lq infimum\rq~should be interpreted as \lq minimum\rq. The \emph{min-plus deconvolution} of functions $f, g\in\mathcal{F}$, denoted by $\oslash$, is defined as  
  \begin{displaymath}
     (f\oslash g)(t) = \sup_{ s\ge 0}\{f(s+t)-g(s)\}
  \end{displaymath}
where, when it applies,  \lq supremum\rq~should be interpreted as \lq maximum\rq.

In the max-plus algebra, the \lq addition\rq~operation represents \emph{supremum} or \emph{maximum} when it exists, and the \lq multiplication\rq~operation is $+$. The \emph{max-plus convolution} of functions $f, g\in\mathcal{F}$, denoted by $\bar{\otimes}$, is defined as  
  \begin{displaymath}
     (f\bar{\otimes} g)(n) = \sup_{0\le m\le n}\{f(m)+g(n-m)\}
  \end{displaymath}
where, when it applies, \lq supremum\rq~should be interpreted as \lq maximum\rq. The \emph{max-plus deconvolution} of functions $f, g\in\mathcal{F}$, denoted by $\bar{\oslash}$, is defined as  
  \begin{displaymath}
     (f\bar{\oslash} g)(n) = \inf_{m\geq 0}\{f(n + m) - g(m)\}
  \end{displaymath}
where, when it applies,  \lq supremum\rq~should be interpreted as \lq maximum\rq.

The max-plus convolution is associative and commutative \cite{Boudec:DeterNC}.  
\begin{itemize}
  \item Associativity: for any $g_1,g_2,g_3\in\mathcal{F}$, $(g_1\bar{\otimes} g_2)\bar{\otimes} g_3 = g_1\bar{\otimes} (g_2\bar{\otimes} g_3)$.
  \item Commutativity: for any $g_1,g_2\in\mathcal{F}$, $g_1\bar{\otimes} g_2 = g_2\bar{\otimes} g_1$. 
\end{itemize}

\subsection{State of The Art in Stochastic Network Calculus}\label{Subsec-stateofart}

The available literature on stochastic network calculus mainly focuses on modeling network behavior and analyzing network performance from the spatial perspective \cite{Burchard:MinPlus}\cite{Ciucu:Scaling}\cite{Fidler:EndMGF}\cite{Fidler:Survey}\cite{Jiang:BasicStoNet}\cite{Jiang:book}\cite{Jiang:Fundamental}\cite{Li:CalculusEB}\cite{Liu:ACalculus}\cite{Mao:Survey}. We call the corresponding models and results \emph{space-domain} models and results in this paper.

In order to characterize the arrival process of a flow from the spatial perspective, let us consider the amount of traffic generated by this flow in a time interval $(s,t]$, denoted by $\mathcal{A}(s,t)$. In the context of stochastic network calculus, the arrival curve model is defined based on a stochastic upper bound on the cumulative amount of the arrival traffic. Here, we only review one relevant space-domain arrival curve model, \emph{virtual-backlog-centric} ($v.b.c$) stochastic arrival curve (SAC) \cite{Jiang:book}.  
\begin{definition}\label{vbaArrCur}
\textbf{(v.b.c Stochastic Arrival Curve)} 

A flow is said to have a virtual-backlog-centric (v.b.c) stochastic arrival curve $\alpha(t)\in\mathcal{F}$ with bounding function $f(x)\in\bar{\mathcal{F}}$, if for all $0\le s\le t$ and all $x\ge 0$, there holds
\begin{eqnarray}
P\Big\{\sup_{0\le s\le t}\big[\mathcal{A}(s,t) - \alpha(t-s)\big]> x\Big\}\le f(x).\label{eq:vbcArrCur}
\end{eqnarray} 
\end{definition}

In stochastic network calculus, the service curve model is defined as a stochastic lower bound on the cumulative amount of service provided by the system. Two space-domain service curve models \cite{Jiang:book} are reviewed here.

\begin{definition}\label{WeakSerCur}
\textbf{(Weak Stochastic Service Curve)} 

A network system is said to provide a weak stochastic service curve $\beta(t)\in\mathcal{F}$ with bounding function $g(x)\in\bar{\mathcal{F}}$ for the arrival process $\mathcal{A}(t)$, if for all $t\ge 0$ and all $x\ge 0$, there holds
\begin{eqnarray}
P\big\{\mathcal{A}\otimes\beta(t)-\mathcal{A}^*(t)>x\big\}\le g(x), \label{eq:WeakSerCur}
\end{eqnarray}
where $\mathcal{A}^*(t)$ denotes the cumulative amount of the departure traffic. 
\end{definition}

Unlike the arrival curve, it is difficult to identify the service curve from (\ref{eq:WeakSerCur}) because it couples the arrival process, the service curve and the departure process. Thus we need a more explicit model to directly reveal the relation between the service process and its service curve such as the following model \cite{Jiang:BasicStoNet}. 

\begin{definition}\label{StrictSerCur}
\textbf{(Stochastic Strict Service Curve)} 

A network system is said to provide stochastic strict service curve $\beta(t)\in\mathcal{F}$ with bounding function $g(x)\in\bar{\mathcal{F}}$, if during any period $(s,t]$, the amount of service $\mathcal{S}(s,t)$ provided by this system satisfies, for any $x\ge 0$,  
\begin{eqnarray}
P\big\{\mathcal{S}(s,t)<\beta(t-s)-x\big\}\le g(x). \label{eq:SpaceStrcitSerCur}
\end{eqnarray}
\end{definition}
Definition \ref{StrictSerCur} is applied to \lq any period\rq~which implies both worst-case scenario and other scenarios. If we could determine a function $\beta(t)$ which makes Eq.(\ref{eq:SpaceStrcitSerCur}) hold under the worst-case scenario, then Eq.(\ref{eq:SpaceStrcitSerCur}) automatically holds under other scenarios as well.       

Based on the arrival curve and service curve models, five fundamental properties have been proved to facilitate tractable analysis. For example, they can be used to derive service guarantees including delay bound and backlog bound, characterize the behavior of traffic departing from a server, describe the service provided along a multi-node path, determine the arrival curve for the aggregate flow, and compute the service provided to each constituent flow. 

\newpage
\begin{itemize}
\item \textbf{P.1: Service Guarantees (single-node)}

Under the condition that the traffic arrival process has an arrival curve $\alpha(t)$ with bounding function $f(x)$ and the network node provides service with a service curve model $\beta(t)$ with bounding function $g(x)$, the stochastic delay bound and stochastic backlog bound can be derived. Particularly, the backlog bound is related to the \emph{maximal vertical distance} between $\alpha(t)$ and $\beta(t)$; the delay bound is relevant to the \emph{maximal horizontal distance} between $\alpha(t)$ and $\beta(t)$. 
    
\item \textbf{P.2: Output Characterization} 

To analyze the end-to-end performance of a multi-hop path, one option is the node-by-node analysis approach. This approach requires being able to characterize the traffic behavior after the traffic has been served and leaves the previous node. The output process of a flow from a node can also be characterized by an arrival curve which is determined by both the arrival curve of the arrival process and the service curve of the service process.     

\item \textbf{P.3: Concatenation Property (multi-node)}

Network calculus possesses an unique property, concatenation property, which is also used to analyze the end-to-end performance but improves the results obtained from the node-by-node analysis. The essence of the concatenation property is to represent a series of nodes in tandem as a \lq black box\rq~which can be treated as a single node. The service curve of this equivalent system is determined by the service curve of all individual nodes along this path. 

\item \textbf{P.4: Superposition Property (aggregate flow)}

Flow aggregation is very common in packet-switched networks. If multiple flows are aggregated into a single flow under the FIFO order, the aggregate flow also has an arrival curve which is the summation of the arrival curve of all constituent flows. 

\item \textbf{P.5: Leftover Service Characterization (per-flow)}

The leftover service characterization makes per-flow performance analysis feasible under FIFO aggregate scheduling. The crucial concept is to represent all other constituent flows as an \lq aggregate cross flow\rq~which can be characterized using an arrival curve. Then the service provided to the constituent flow of interest can also be described by a service curve which is determined by the service curve provided to all arrival flows and the arrival curve of the \lq aggregate cross flow\rq. 

\end{itemize}

The superposition property of the $v.b.c$ SAC \cite{Jiang:book} is reviewed here because it is relevant to the model transformation in the following content.  

\begin{theorem}\label{ThSpaceSuperposition}

Consider $N$ flows with arrival processes $\mathcal{A}_i$, $i=1,...,N$, respectively.\nop{ Let $\mathcal{A}$ denote the aggregate arrival process.} If each arrival process has a $v.b.c$ SAC $\alpha_i(t)\in\mathcal{F}$ with bounding function $f_i(x)\in\bar{\mathcal{F}}$, then the aggregate arrival process has a $v.b.c$ SAC $\alpha(t)\in\mathcal{F}$ with bounding function $f(x)\in\bar{\mathcal{F}}$, where
\begin{eqnarray}
\alpha(t) &=& \sum_{i=1}^N \alpha_i(t)~~ \text{and} ~~f(x) = f_1\otimes\cdot\cdot\cdot\otimes f_N(x).\nonumber
\end{eqnarray}
\end{theorem}

\section{Time-domain Modeling and Transformations}\label{Sec-ModelDefinition}

This section defines traffic and service models in the \emph{time-domain}. Particularly, traffic models are defined based on probabilistic lower bounds on the \emph{cumulative inter-arrival time} between two arbitrary packets. Service models are defined in terms of the virtual time function and probabilistic upper bounds on the \emph{cumulative service time} between two arbitrary packets. Moreover, we establish the transformations among these models as well as the transformation between the time-domain model and the space-domain model. 

\subsection{Time-domain Traffic Models}\label{Section-Model}
Consider an arrival process that specifies packets arriving to a network system at time $a(n)$, $n=0,1,2,...$. In order to stochastically guarantee a certain level of QoS to this arrival process, this arrival process should be constrained. By characterizing the constrained arrival traffic from the temporal perspective, we define an inter-arrival-time (i.a.t) stochastic arrival curve model.  

\begin{definition}\label{itarrcurve}
\textbf{(i.a.t Stochastic Arrival Curve)} 

A flow is said to have an inter-arrival-time (i.a.t) stochastic arrival curve $\lambda(n)\in\mathcal{F}$ with bounding function $h(x)\in\bar{\mathcal{F}}$, if for any $m,n \geq 0$ and $x\geq 0$, there holds
\begin{equation}
P\Big\{a(m+n)-a(n) < \big[\lambda(n) - x\big]^+\Big\}\leq h(x).
\label{eq:itarricurve}
\end{equation}
\nop{where $[z]^+=\max[z,0]$.}
\end{definition}
Eq.(\ref{eq:itarricurve}) indicates that function $\lambda(n)$ is a probabilistic lower bound on the cumulative inter-arrival time. The violation probability that the cumulative inter-arrival time is smaller than $\lambda(n)$ is bounded above by function $h(x)$. If $h(x)=0$ for all $x\ge 0$, Eq.(\ref{eq:itarricurve}) represents a time-domain deterministic arrival curve \cite{Chang:MaxPlus} which is a special case of the $i.a.t$ SAC.       

Queueing theory typically characterizes the arrival process using the probability distribution of the inter-arrival time between two consecutive customers: $$P\{a(n)-a(n-1) \leq x\}=F(x).$$ Comparing $F(x)$ with Eq.(\ref{eq:itarricurve}), we notice that Eq.(\ref{eq:itarricurve}) gives a more general probability expression of the inter-arrival time between two arbitrary packets. From this viewpoint, $F(x)$ is a special case of Eq.(\ref{eq:itarricurve}).  

\begin{example}\label{exampleidarricurve}
\end{example} Consider a flow of packets with fixed packet size. Suppose that packet inter-arrival times follow an exponential distribution with mean $1/\mu$. Then, the packet arrival time has an Erlang distribution with parameter $(n,\mu)$ \cite{handbook:erlangdist}, where $n$ denotes the number of arrival packets. For any two packets $P(m)$ and $P(m+n)$, their inter-arrival time satisfies, for $x\geq 0$,
\begin{eqnarray}
P\Big\{a(m+n)-a(m) < \frac{n}{\mu} - x\Big\}&\leq& P\Big\{a(m+n)-a(m) \leq \big[\frac{n}{\mu} - x\big]^+\Big\} \nonumber\\
&=& 1-\sum_{k=0}^{n-1}\frac{e^{-\mu y}(\mu y)^k}{k!}\nonumber
\end{eqnarray}
where $y=\frac{n}{\mu}-x$. Thus, the flow has an $i.a.t$ SAC $\lambda(n)=\frac{n}{\mu}$. 

The $i.a.t$ SAC is simple but has limited applications. For example, consider a virtual single server queue (SSQ) fed with the arrival traffic which has an $i.a.t$ SAC $\lambda(n)$ with bounding function $h(x)$. Suppose that the virtual SSQ provides a constant service time $\lambda(1)$ for each packet. From Eq.(\ref{eq:queueingdelay}), the waiting delay of $P(n)$ experienced in the virtual SSQ is 
\begin{eqnarray}
W(n)&=&d(n)-a(n)-\lambda(1)\nonumber\\
&=& \sup_{0\le m\le n}\big[a(m)+\lambda(n-m+1)\big]-a(n)-\lambda(1)\label{stepdeparturetime}\\
&=& \sup_{0\le m\le n}\Big\{\lambda(n-m)-\big[a(n)-a(m)\big]\Big\}\label{eq:difficulty2}
\end{eqnarray}
where $a(m)$ is the beginning of the backlogged period within which packet $P(n)$ is transmitted. Eq.(\ref{stepdeparturetime}) is derived from the departure time given in Eq.(\ref{eq:sumservicetimedeparture}). Eq.(\ref{eq:difficulty2}) is called the \emph{virtual-waiting-delay} property. It is difficult to compute the virtual-waiting-delay from Eq.(\ref{eq:itarricurve}). When investigating the performance guarantees such as delay bound and backlog bound in Section \ref{Sec-serviceguarantee}, we face the similar difficulty. 

In order to deal with the difficulty of computing the virtual-waiting-delay, we define another stochastic arrival curve model based on Eq.(\ref{eq:difficulty2}).  

\begin{definition}\label{vstarrcurve}
\textbf{(v.w.d Stochastic Arrival Curve)} 

A flow is said to have a virtual-waiting-delay (v.w.d) stochastic arrival curve $\lambda(n)\in\mathcal{F}$ with bounding function $h(x)\in\bar{\mathcal{F}}$, if for any $0\leq m\leq n$ and $x\geq 0$, there holds
\begin{equation}
P\Big\{\sup_{0\leq m\leq n}\big\{\lambda(n-m) - \big[a(n)-a(m)\big]\big\} > x\Big\}\leq h(x).
\label{eq:vsdArrCurveNew}
\end{equation}
\end{definition}
Through some manipulations, Eq.(\ref{eq:vsdArrCurveNew}) can be expressed as the max-plus convolution:
\begin{eqnarray}
P\big\{a(n) < a\bar{\otimes}\lambda(n) - x\big\}\leq h(x).
\label{eq:vsdArrCurve1}
\end{eqnarray}
Here, $a\bar{\otimes}\lambda(n)$ can be considered as the expected time that the packet would arrive to the head-of-line (HOL) if the flow has passed through a virtual SSQ with the (deterministic) service curve $\lambda(n)$. The packet is expected to arrive not earlier than the expected HOL time. Here $x$ represents the difference between the expected HOL time and the actual arrival time. The violation probability is bounded by the non-increasing function $h(x)$.  

We use the $v.w.d$ SAC to characterize the arrival traffic in Example \ref{exampleidarricurve}. 
\begin{example}\label{examWait}
\end{example} Consider a flow that consists of packets having the fixed packet size. Suppose that all packet inter-arrival times are exponentially distributed with mean 
$\frac{1}{\mu}$. Based on the steady-state probability mass function (PMF) of the queue-waiting time for an M/D/1 queue \cite{Shortle:MD1queue}, we say that the flow has a $v.w.d$ SAC $\lambda(n)=\hbar\cdot n$ with bounding function $h^{exp}$ for $0<\hbar<\frac{1}{\mu}$. Let $\rho=\mu\cdot\hbar$. We obtain the bounding function of the probability that the waiting delay $W(n)$ exceeds $x(\geq 0)$ 
\begin{displaymath}
h^{exp}(x) = 1 - (1-\rho)\sum_{i=0}^{\lfloor\frac{x}{\hbar}\rfloor}e^{-\mu(i\hbar-x)}\frac{[\mu(i\hbar-x)]^i}{i!}
\end{displaymath}
where, $\lfloor y\rfloor$ denotes the floor function. 

The definition of $v.w.d$ SAC is more strict than that of $i.a.t$ SAC. As a result, it is not trivial to derive the $v.w.d$ SAC for an arrival process even if it can be characterized by an $i.a.t$ SAC. Thus, it is important to explore whether there exists some relationship between the $i.a.t$ SAC and the $v.w.d$ SAC.

\begin{theorem}\label{it2vsdrelation}
\begin{enumerate}
\item If a flow has a v.w.d SAC $\lambda(n)\in\mathcal{F}$ with bounding function $h(x)\in\bar{\mathcal{F}}$, then the flow has an i.a.t SAC $\lambda(n)\in\mathcal{F}$ with the same bounding function $h(x)\in\bar{\mathcal{F}}$.
\item Conversely, if a flow has an i.a.t SAC $\lambda(n)\in\mathcal{F}$ with bounding function $h(x)\in\bar{\mathcal{G}}$, it also has a v.w.d SAC $\lambda_{-\eta}(n)\in\mathcal{F}$ with bounding function $h_{\eta}(x)\in\bar{\mathcal{G}}$, where for $\eta > 0\footnote{Note that $\eta$ should not be greater than $\lim_{n\rightarrow\infty}\frac{\lambda(n)}{n}$.}$ 
\begin{eqnarray}
\lambda_{-\eta}(n) = [\lambda(n) - \eta\cdot n]^+ ~~\text{and} ~~ h_{\eta}(x) &=& \Big[h(x)+\frac{1}{\eta}\int_{x}^{\infty}h(y)dy\Big]_1.\nonumber
\end{eqnarray}

\end{enumerate}
\end{theorem}  
\textbf{Remark.} In the second part, $h(x)\in\bar{\mathcal{G}}$ while not $h(x)\in\bar{\mathcal{F}}$. If the requirement on the bounding function is relaxed to $h(x)\in\bar{\mathcal{F}}$, the second part may not hold in general. 

Theorem \ref{it2vsdrelation} reveals that if an arrival process can be modeled by a $v.w.d$ SAC $\lambda(n)$, then $\lambda(n)$ is also the $i.a.t$ SAC of this arrival process. On the other hand, if an arrival process can be modeled by an $i.a.t$ SAC with the associated bounding function in $\bar{\mathcal{G}}$, then this arrival process also has a $v.w.d$ SAC which may be associated with a more loose bounding function. 

It is worth highlighting that the $v.w.d$ SAC looks similar to the $v.b.c$ SAC (see Definition \ref{vbaArrCur}) defined in the space-domain. Since these two models play an important role in performance analysis in their respective domains, we establish their relationship in the following theorem. 
 
\begin{theorem}\label{vbctovsd}
\begin{enumerate}
\item If a flow has a space-domain v.b.c SAC $\alpha(t)\in\mathcal{F}$ with bounding function $f(x)\in\bar{\mathcal{F}}$, the flow has a time-domain v.w.d SAC $\lambda(n)\in\mathcal{F}$ with bounding function $h(y)\in\bar{\mathcal{F}}$, where
\begin{displaymath}
 \lambda(n)=\inf\{\tau:\alpha(\tau)\geq n\},~~ \text{and}~~ h(y) = f\big(z^{-1}(y)\big)
\end{displaymath} 
with $z^{-1}(y)$ denoting the inverse function of $y$, where $$y= z(x) \equiv \sup_{k\ge 0} \{\lambda(k)-\lambda(k-x)\}.$$ Specifically, if $\lambda(\cdot)$ is sub-additive, $z(x) = \lambda(x)$.

\item Conversely, if a flow has a time-domain v.w.d SAC $\lambda(n)\in\mathcal{F}$ with bounding function $h(y)\in\bar{\mathcal{F}}$, the flow has a space-domain v.b.c SAC $\alpha(t)\in\mathcal{F}$ with bounding function $f(x)\in\bar{\mathcal{F}}$, where 
\begin{displaymath}
\alpha(t)=\sup\{k:\lambda(k)\leq t\},~~ \text{and} ~~f(x) = h\big(z^{-1}(x)\big)
\end{displaymath}
with $z^{-1}(x)$ denoting the inverse function of $x$, where $$x= z(y) \equiv \sup_{\tau \ge 0} \{ \alpha(\tau+y) - \alpha(\tau) +1\}.$$ Specifically, if $\alpha(\cdot)$ is sub-additive\footnote{\cite{Boudec:AppNC} clarifies that $\alpha(t)$ defines a meaningful constraint only if it is subadditive. If $\alpha(t)$ is not subadditive, it can be replaced by its subadditive closure.}, $z(y) = \alpha(y)+1$. 
\end{enumerate}
\end{theorem} 
\textbf{Note} that in Theorem \ref{vbctovsd}, the arrival curve $\alpha(t)$ denotes the cumulative number of arrival packets while not the cumulative amount (in bits) of arrival traffic.  

The generalized stochastically bounded burstiness (gSBB) \cite{Yin:gSBB} is a special case of the space-domain $v.b.c$ SAC. A summarization of some well-known traffic belonging to gSBB is given \cite{Jiang:book}, including both Gaussian self-similar processes \cite{Addie:Gaussian}\cite{Cheo:Gaussian}\cite{Kim:Gaussian}\cite{Mannersalo:Gaussian}, such as fractional Brownian motion, and non-Gaussian self-similar processes, such as $\alpha-$stable self-similar process \cite{AK:Self}\cite{Karasaridis:NonGaussian}, and the $(\sigma(\theta),\rho(\theta))$ stochastic traffic model \cite{Change:traffic}\cite{Chang:DeterNC}. With Theorem \ref{vbctovsd}, the following example shows that gSBB can be readily represented using the time-domain $v.w.d$ SAC.   

\begin{example}\label{eggsBB}
\end{example} If an arrival process $\mathcal{A}(t)$ can be described by gSBB with upper rate $\rho$ and bounding function $f(x)\in\bar{\mathcal{F}}$, i.e., for any $t, x\geq 0$, there holds 
\begin{displaymath}
P\Big\{\sup_{0\leq s\leq t}\big\{\mathcal{A}(s,t)-\rho\cdot(t-s)\big\}> x \Big\}\leq f(x),
\end{displaymath}
then the process $\mathcal{A}(t)$ has a $v.b.c$ SAC $\alpha(t)=\rho\cdot t$ with the bounding function $f(x)$. With Theorem \ref{vbctovsd}~(1), the arrival process has a $v.w.d$ SAC $\lambda(n)=\frac{n}{\rho}$ which is sub-additive and the bounding function $h(y)=f(\rho \cdot y)$, i.e.,  
\begin{displaymath}
P\Big\{\sup_{0\leq m \leq n}\big\{\frac{1}{\rho}\cdot (n-m)- [a(n) - a(m)]\big\} > y \Big\} \leq f(\rho\cdot y).
\end{displaymath}

\textbf{Remark.} Theorem \ref{vbctovsd} allows us to readily utilize the results of gSBB traffic for time-domain models. If the traffic is more suitable for being characterized by the time-domain traffic models rather than the space-domain traffic models, then the transformation between two domains can facilitate the analysis.     

\subsection{Time-domain Service Models}\label{Sec-server}
Queueing theory characterizes the service process of a system based on the per customer service time. Like the arrival model, time-domain service models extend to the cumulative service time.  

If packet $P(n)$ arrives to a network system after packet $P(n-1)$ has departed from the system, the departure time of $P(n)$ is the arrival time $a(n)$ plus the service time $\delta_n$, i.e., $a(n)+\delta_n$. If $P(n)$ arrives to the system while $P(n-1)$ is still in the system, then its departure time is $d(n-1)+\delta_n$. The combination of both cases gives the departure time of $P(n)$ 
\begin{equation}
d(n) = \max[a(n),d(n-1)] + \delta_n
\label{eq:departtime}
\end{equation}
with $d(0)=a(0)+\delta_0$. Applying Eq.(\ref{eq:departtime}) iteratively to its right-hand side results in 
\begin{equation}
d(n) = \sup_{0\leq m \leq n}\big[a(m) + \sum_{k=m}^n\delta_k\big].
\label{eq:sumservicetimedeparture}
\end{equation}
The system usually allocates a minimum service rate to an arrival flow in order to meet its QoS requirements. The guaranteed minimum service rate is related to the guaranteed maximum service time for each packet of the flow. Accordingly, the time that the packet departs from the system is bounded. Denote the guaranteed maximum service time by $\hat{\delta}_n$. The \emph{Guaranteed Rate Clock} (GRC) is defined based on $\hat{\delta}_n$ \cite{Goyal:GR}~\cite{PGoyal:Generalized}: 
\begin{eqnarray}
GRC(n) = \max[a(n),GRC(n-1)]+\hat{\delta}_n.\label{eq:GRClock}
\end{eqnarray}
with $GRC(0)=a(0)+\hat{\delta}_0$. Applying Eq.(\ref{eq:GRClock}) iteratively to its right-hand side yields 
\begin{equation}
GRC(n) = \sup_{0\leq m \leq n}\big[a(m) + \sum_{k=m}^n\hat{\delta}_k\big]. ~~~~
\label{eq:departuretimeori}
\end{equation}

Eq.(\ref{eq:departuretimeori}) is similar to Eq.(\ref{eq:sumservicetimedeparture}) except for that $GRC(n)$ represents the guaranteed departure time\footnote{The guaranteed departure time is actually $GRC(n)+$error term \cite{Goyal:GR}, where \emph{error term} is determined by the employed service discipline. The underlying service discipline considered throughout this paper is FIFO, under which, the error term is $zero$.} while $d(n)$ is the actual departure time. 

If $\sum_{i=m}^n\hat{\delta}_i$ is denoted by a function $\gamma(n-m+1)$, then Eq.(\ref{eq:departuretimeori}) becomes
\begin{eqnarray}
GRC(n) = \sup_{0\leq m \leq n}\big[a(m) + \gamma(n-m+1)\big] = a\bar{\otimes}\gamma(n)\label{eq:ExpDepartTime}
\end{eqnarray}
which is the basis for the time-domain (deterministic) service model \cite{Chang:MaxPlus}. For systems that only provide service guarantees stochastically or applications that require only stochastic QoS guarantees, the service time may not need to be deterministically guaranteed. In this case, we extend the (deterministic) service curve into a probabilistic one.

\begin{definition}\label{StoSerCurve}
\textbf{(i.d Stochastic Service Curve)} 

A system is said to provide an \emph{inter-departure time (i.d) stochastic service curve} $\gamma(n)\in\mathcal{F}$ with bounding function $j(x)\in\bar{\mathcal{F}}$, if for any $n, x\geq 0$, there holds
\begin{equation}
P\Big\{d(n) - a\bar{\otimes}\gamma(n) > x\Big\}\leq j(x).
\label{eq:idservicecurve1}
\end{equation} 
\end{definition}
Note that the stochastic service curve of a service process is not unique. Therefore optimization is needed to find the SSC of a specific system. 

\begin{example}\label{eg-servicecurve}
\end{example} Consider two nodes, the transmitter and the receiver. They communicate through an error-prone wireless link which is modeled as a slotted system. The wireless link can be considered as a stochastic server. Packets have fixed-length and are served in a FIFO manner by the transmitter. To simplify the analysis, we assume that the length of time slot equals one packet transmission time\footnote{It means we only compute the number of time slots in this example.}. 

The transmitter sends packets only at the beginning of a time slot. Due to the error-prone nature of the wireless link, the probability that a packet is successfully transmitted is determined by packet error rate (PER). Here, we assume that packet errors happen independently in every transmission with a fixed PER denoted by $P_e$. The successful transmission probability of one packet is hence $1-P_e$. If error happens, the unsuccessfully transmitted packet will be retransmitted in the next time slot immediately. In order to guarantee 100\% reliability, the packet will be retransmitted until it is successfully received by the receiver. 

The per-packet service time $\delta_n$ is a geometric random variable with parameter $1-P_e$. The cumulative service time of successfully transmitting 
packets $P(m)$ to $P(n)$ is $\sum_{k=m}^n\delta_k$ which follows the negative binomial distribution with parameter $1-P_e$. The mean service time denoted by $\bar{\delta}$ equals $\frac{1}{1-P_e}.$ 

According to the complementary cumulative distribution function (CCDF) of the negative binomial distribution, the cumulative service time between two arbitrary packets $P(m)$ and $P(m+n)$ is given by
\begin{eqnarray}
P\Big\{\sum_{k=m}^{m+n}\delta_k >\bar{\delta}\cdot(n+1) + x\Big\}\leq \sum_{i=\lceil\gamma(n+1)+x\rceil}^{\infty}\begin{pmatrix}i-1\\n\end{pmatrix}(1-P_e)^{n+1}P_e^{i-(n+1)}
\label{eq:wirelesslinkserbounding}
\end{eqnarray}
for any $x\ge 0$, where $\lceil\cdot\rceil$ is the ceiling function. 

The right-hand side of Eq.(\ref{eq:wirelesslinkserbounding}) represents the bound on the probability that the actual cumulative service time exceeds the cumulative mean service time. Let $\gamma_{\eta}(n)=\bar{\delta}\cdot n+\eta\cdot n$ for $\eta>0$ and $j(x)$ denote the right-hand side of Eq.(\ref{eq:wirelesslinkserbounding}). From Definition \ref{StoSerCurve}, we know
\begin{eqnarray}
&&d(n) - a\bar{\otimes}\gamma_{\eta}(n) \nonumber\\
&=& \sup_{0\leq m \leq n}\big[a(m)+\sum_{k=m}^n\delta_k\big]-\sup_{0\leq m \leq n}\big[a(m)+(\bar{\delta}+\eta)\cdot(n-m+1)\big]\nonumber\\
&\leq& \sup_{0\leq m \leq n}\big[\sum_{k=m}^{n}\delta_k - \bar{\delta}\cdot(n-m+1)-\eta\cdot(n-m+1)\big],\nonumber
\end{eqnarray}
from which, we have 
\begin{eqnarray}
&&P\Big\{\sup_{0\leq m \leq n}\big[\sum_{k=m}^{n}\delta_k - \bar{\delta}\cdot(n-m+1)-\eta\cdot(n-m+1)\big]> x\Big\}\nonumber\\
&\leq& \sum_{m=0}^{n}P\Big\{\sum_{k=m}^{n}\delta_k - \bar{\delta}\cdot(n-m+1)> x+\eta\cdot(n-m+1)\Big\}\nonumber\\
&\leq& \sum_{m=0}^n j(x+\eta\cdot(n-m+1))\nonumber
\end{eqnarray}
\begin{eqnarray}
&=&\sum_{k=1}^{n+1} j(x+\eta\cdot k) \leq \Big[\frac{1}{\eta}\int_{x}^{\infty}j(y)dy\Big]_1.\nonumber
\end{eqnarray}
Thus, we conclude that this error-prone wireless link provides an $i.d$ SSC $\gamma_\eta(n)$ with the bounding function $j_\eta(x)$ for $\eta>0$, where
\begin{eqnarray}
\gamma_{\eta}(n)=\bar{\delta}\cdot n+\eta\cdot n ~~\text{and}~~
j_{\eta}(x) = \Big[\frac{1}{\eta}\int_{x}^{\infty}j(y)dy\Big]_1.\nonumber
\end{eqnarray}

Since Eq.(\ref{eq:wirelesslinkserbounding}) is only relevant to the cumulative service time and does not involve the arrival process, it provides a method to find the $i.d$ SSC. 

\textbf{Remark.} Example \ref{eg-servicecurve} demonstrates that we can obtain the $i.d$ SSC from analyzing per-packet service time. However, if applying the space-domain results to this case, we need an impairment process \cite{Jiang:book} to characterize the cumulative amount of service consumed by unsuccessful transmissions. In other words, we still need to compute the cumulative slots due to failed transmission and then convert it into the amount of service. Such conversion may introduce error or result in looser bounds whereas the time-domain model directly computes the service time and avoids the conversion error. This simple example thus illustrates the feasibility of the time-domain service curve model.     

In Section \ref{Sec-Property}, we show that many results can be derived from the $i.d$ SSC. However, without additional constraints, we have difficulty in proving the concatenation property for the $i.d$ SSC. To address this difficulty, we introduce another service curve model in the following. 

\begin{definition}\label{ConstrainedSerCurve}
\textbf{($\eta$-Stochastic Service Curve)}

A system is said to provide an \emph{$\eta$-stochastic service curve} $\gamma(n)\in\mathcal{F}$ with bounding function $j_{\eta}(x)\in\bar{\mathcal{F}}$, if for any $n, x\geq 0$, there holds
\begin{equation}
P\Big\{\sup_{0\leq m\leq n}\big[d(m) - a\bar{\otimes}\gamma(m) - \eta\cdot(n-m)\big] > x\Big\}\leq j^{\eta}(x),\label{eq:etaService}
\end{equation} 
for any small $\eta>0$. 
\end{definition}
Note that the left-hand side of Eq.(\ref{eq:etaService}) represents a property that is typically hard to calculate. It means that Definition \ref{ConstrainedSerCurve} is more strict than Definition \ref{StoSerCurve}. Thus it is important to find the relationship between the $i.d$ SSC and the $\eta$-stochastic service curves. 

\begin{theorem}\label{stosercur2constosercur}
\begin{enumerate}
\item If a system provides to its arrival process an $\eta$-stochastic service curve $\gamma(n)$ with bounding function $j_{\eta}(x)\in\bar{\mathcal{F}}$, it provides to the arrival process an $i.d$ SSC $\gamma(n)$ with the same bounding function $j_{\eta}(x)\in\bar{\mathcal{F}}$;

\item If a system provides to its arrival process an $i.d$ SSC $\gamma(n)$ with bounding function $j(x)\in\bar{\mathcal{G}}$, it provides to the arrival process an $\eta$-stochastic service curve $\gamma(n)$ with bounding function $j_{\eta}(x)\in\bar{\mathcal{G}}$ for $\eta>0$, where 
\begin{displaymath}
j_{\eta}(x) = \Big[j(x)+\frac{1}{\eta}\int_{x}^{\infty} j(y)dy\Big]_1.
\end{displaymath}
\end{enumerate}
\end{theorem}
Again, in the second part of Theorem \ref{stosercur2constosercur}, $j(x)\in\bar{\mathcal{G}}$ while not $j(x)\in\bar{\mathcal{F}}$. If the requirement on the bounding function is relaxed to $j(x)\in\bar{\mathcal{F}}$, the above relationship may not hold in general. 

Definition \ref{StoSerCurve} explores the relationship between the arrival process and the departure process, but it does not explicitly characterize the service process. From Eq.(\ref{eq:idservicecurve1}), it is not trivial to find the stochastic service curve $\gamma(n)$ for a specific system. Example \ref{eg-servicecurve} illustrates how to add some increment $\eta$ to the stochastic service curve $\gamma(n)$. To this end, we expand Eq.(\ref{eq:ExpDepartTime}) to
\begin{equation}
d(n)-a\bar{\otimes}\gamma(n) = \sup_{0\le m\le n}\big[a(m)+\Delta(m,n)\big]-a\bar{\otimes}\gamma(n).\label{eq:serviceinterstep1}
\end{equation}
\nop{Recall $\Delta(m,n)=\sum_{k=m}^n\delta_k$. }Without loss of generality, assume $a(m_0)$ ($0\le m_0\le n$) is the beginning of the backlogged period in which packet $P(n)$ is served. Then, 
$$\sup_{0\le m\le n}\big[a(m)+\Delta(m,n)\big]=a(m_0)+\Delta(m_0,n)$$ 
and $a\bar{\otimes}\gamma(n)\ge a(m_0)+\gamma(n-m_0+1)$. 

We rewrite the right-hand side of Eq.(\ref{eq:serviceinterstep1}) as 
\begin{eqnarray} 
&&a(m_0)+\Delta(m_0,n)-a\bar{\otimes}\gamma(n)\nonumber\\
&\le& a(m_0)+\Delta(m_0,n) - a(m_0)-\gamma(m_0+1)\nonumber\\
&=&\Delta(m_0,n)-\gamma(n-m_0+1).\label{eq:relationstocservice2strict}
\end{eqnarray}
Note that Eq.(\ref{eq:relationstocservice2strict}) holds for arbitrary $m_0\leq n$. Inspired by this, we define a new service curve model.

 \newpage
\begin{definition}\label{strictservicecurve}
\textbf{(Stochastic Strict Service Curve)}

A system is said to provide stochastic strict service curve $\gamma(n)\in\mathcal{F}$ with bounding function $j(x)\in\bar{\mathcal{F}}$, if the cumulative service time between two arbitrary packets $P(m)$ and $P(n)$\footnote{If $P(m)$ and $P(n)$ are in the same backlogged period, $\Delta(m,n)=d(n)-d(m-1)$.} satisfies for any $x\ge 0$, 
\begin{eqnarray}
&& P\Big\{\Delta(m,n)-\gamma(n-m+1)>x\Big\}\le j(x).\label{eq:DefStrictService}
\end{eqnarray}
\end{definition}

Eq.(\ref{eq:relationstocservice2strict}) reveals a relationship between the $i.d$ SSC and the stochastic strict service curve. Furthermore, in Theorem \ref{stosercur2constosercur}(2), the relationship between the stochastic strict service curve and the $\eta-$stochastic service curve is obtained.  
\begin{theorem}\label{id2stostrict}
Consider a system providing stochastic strict service curve $\gamma(n)\in\mathcal{F}$ with bounding function $j(x)\in\bar{\mathcal{F}}$. 
\begin{enumerate}
  \item It provides an $i.d$ SSC $\gamma(n)$ with the same bounding function $j(x)$.
  \item If $j(x)\in\bar{\mathcal{G}}$, it provides an $\eta-$stochastic service curve $\gamma(n)$ with bounding function $j_{\eta}(x)\in\bar{\mathcal{G}}$, where 
\begin{displaymath}
j_{\eta}(x) = \Big[j(x)+\frac{1}{\eta}\int_{x}^{\infty} j(y)dy\Big]_1.
\end{displaymath}
\end{enumerate}
\end{theorem}
 Note that the second part of Theorem \ref{id2stostrict} requires the bounding function $j(x)\in\bar{\mathcal{G}}$ while not $j(x)\in\bar{\mathcal{F}}$.

\section{Fundamental Properties}\label{Sec-Property}
In this section, we explore the four fundamental properties for time-domain models, i.e. service guarantees, output characterization, concatenation property and superposition property. Some properties can only be proved for the combination of a specific traffic model and a specific service mode. This is why we have established various transformations between models in Section \ref{Sec-ModelDefinition}. With these transformations, we can flexibly apply the corresponding models to specific network scenarios. 

\subsection{Service Guarantees}\label{Sec-serviceguarantee}
Suppose that the arrival process has a $v.w.d$ SAC and the service process has an $i.d$ SSC. Under this condition, we derive the delay bound and backlog bound. 

\subsubsection{Delay Bound}\label{Sec-DelayBound}
The system delay significantly impacts QoS and is an important performance metric. 
\begin{theorem}\label{Delaybound}
\textbf{(System Delay Bound).}

Consider that a system provides an i.d SSC $\gamma(n)\in\mathcal{F}$ with bounding function $j(x)\in\bar{\mathcal{F}}$ to the input which has a $v.w.d$ SAC $\lambda(n)\in\mathcal{F}$ with bounding function $h(x)\in\bar{\mathcal{F}}$. Let $D(n) = d(n) - a(n)$ be the system delay of packet $P(n)$. For $x\geq 0$, $D(n)$ is bounded by
\begin{equation}
P\{D(n) > x\} \leq j\otimes h([x - \gamma\oslash\lambda(1)]^+).
\label{eq:DelayBound}
\end{equation} 
\end{theorem}
If the arrival process and the service process are independent of each other, we obtain another system delay bound according to Lemma 6.1 \cite{Jiang:book}. 

\begin{lemma}\label{lemmaIndependentDelay}
\textbf{(System delay bound: independent condition)}

Consider that a system provides an i.d SSC $\gamma(n)\in\mathcal{F}$ with bounding function $j(x)\in\bar{\mathcal{F}}$ to the arrival process which has a $v.w.d$ SAC $\lambda(n)\in\mathcal{F}$ with bounding function $h(x)\in\bar{\mathcal{F}}$. Suppose that the arrival process and the service process are independent of each other. Then for $x\geq 0$, the system delay $D(n)$ is bounded by
\begin{equation}
P\{D(n) > x\} \leq 1-\bar{j}* \bar{h}([x - \gamma\oslash\lambda(1)]^+),
\label{eq:DelayBoundIndependent}
\end{equation} 
where $\bar{j}(x)=1-[j(x)]_1$ and $\bar{h}(x)=1-[h(x)]_1$.  
\end{lemma}

\subsubsection{Backlog Bound}\label{Sec-BacklogBound}
The system backlog represents the total number of packets in the system at time $t$, including both the packets waiting in the buffer and the packet being served. It is determined by function (\ref{eq:timebacklog}):
\begin{displaymath}
B(t) \leq \inf\Big\{l\geq 0, \sup\{n\ge 0: a(n\le t)\}: d(n-l) \leq a(n)\Big\}.
\end{displaymath}
The following theorem provides a probabilistic bound on the system backlog for the given arrival process and service process. 
\begin{theorem}\label{BacklogBound}
\textbf{(Backlog Bound)}

Consider that a system provides an i.d SSC $\gamma(n)\in\mathcal{F}$ with bounding function $j(x)\in\bar{\mathcal{F}}$ to the arrival process which has a $v.w.d$ SAC $\lambda(n)\in\mathcal{F}$ with bounding function $h(x)\in\bar{\mathcal{F}}$. The system backlog at time $t$ ($\geq 0$) is bounded by
\begin{equation}
P\big\{B(t) > x \big\} \leq j\otimes h\big(\gamma\bar{\oslash}\lambda([x-1]^+)\big)
\end{equation}
for $x\geq 1$. 

Let $$H(\lambda,\gamma+x)=\sup_{m\geq 0}\Big\{\inf[k\geq 0:\gamma(m)+x\leq\lambda(m+k)]\Big\}$$ represent the maximum horizontal distance between functions $\lambda(n)$ and $\gamma(n)+x$. The probability that $B(t)$ exceeds $H(\lambda,\gamma+x)$ is bounded by  
\begin{equation}
P\Big\{B(t) > H(\lambda,\gamma+x)+1\Big\} \leq j\otimes h(x).
\label{eq:maxBtbound}
\end{equation}
\end{theorem}

\textbf{Remark.} $H(\lambda,\gamma+x)$ can be considered as the maximum system backlog in a (deterministic) virtual system, where the arrival process is $\lambda(n)$ and the service process is $\gamma(n)+x$. Eq.(\ref{eq:maxBtbound}) is thus a bound on this maximum system backlog.  

If the arrival process and the service process are independent of each other, another backlog bound is derived according to Lemma 6.1 \cite{Jiang:book}. 

\begin{lemma}\label{BacklogBoundIndependent}
\textbf{(Backlog Bound: independent condition)}

Consider that a system provides an i.d SSC $\gamma(n)\in\mathcal{F}$ with bounding function $j(x)\in\bar{\mathcal{F}}$ to the arrival process which has a $v.w.d$ SAC $\lambda(n)\in\mathcal{F}$ with bounding function $h(x)\in\bar{\mathcal{F}}$. Suppose that the arrival process and the service process are independent of each other. Then the system backlog at time $t$ ($\geq 0$) is bounded by:
\begin{equation}
P\big\{B(t) > x \big\} \leq 1-\bar{j}* \bar{h}\big(\gamma\bar{\oslash}\lambda([x-1]^+)\big)\label{eq:BacklogIndependent}
\end{equation}
for $x\geq 1$. 

\nop{Let $H(\lambda,\gamma+x)$ represent the maximum horizontal distance between functions $\lambda(n)$ and $\gamma(n)+x$. }The probability that $B(t)$ exceeds $H(\lambda,\gamma+x)$ is bounded by  
\begin{equation}
P\big\{B(t) > H(\lambda,\gamma+x)+1\big\} \leq 1-\bar{j}* \bar{h}(x).
\label{eq:maxBtboundIndependent}
\end{equation}
\end{lemma}

\subsection{Output Characterization}
The previous section has presented how to derive the service guarantees in a single node. Another common scenario with which performance analysis deals is the end-to-end performance. An intuitive and simple approach is called \emph{node-by-node} analysis \cite{Jiang:DelayBound} which requires characterization of the departure process from a single node.

\begin{figure}[!thbp]
\centering
\includegraphics[width=1\textwidth]{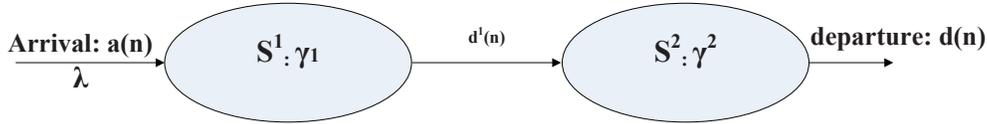}
\caption{Output characterization}
\label{fig:output}
\end{figure}

Let us consider a simple network as shown in Figure \ref{fig:output}. The departure process of Server $1$ is the arrival process for Server $2$.

The delay bound in Server 1 can be derived from the result of Section \ref{Sec-DelayBound}. To derive the delay bound in Server 2, we need to characterize the arrival process to Server 2, which is the departure process from Server 1. The problem is how to characterize the departure process from Server 1. 

\begin{theorem}\label{Outputchar1}
\textbf{(Output Characterization)} 

Consider that a system provides an i.d SSC $\gamma(n)\in\mathcal{F}$ with bounding function $j(x)\in\bar{\mathcal{F}}$ to its arrival process which has a $v.w.d$ SAC $\lambda(n)\in\mathcal{F}$ with bounding function $h(x)\in\bar{\mathcal{F}}$. The output has an i.a.t SAC $\lambda\bar{\oslash}\gamma(n-m-1)$ with bounding function $j\otimes h(x)\in\bar{\mathcal{F}}$, i.e., for any $0\leq m < n-1$, there holds
\begin{equation}
P\Big\{\lambda\bar{\oslash}\gamma(n-m-1) - [d(n) - d(m)] > x\Big\} \leq j\otimes h(x).
\label{eq:outputcharacterization}
\end{equation}
\end{theorem}

\textbf{Remark.} In Theorem \ref{Outputchar1}, the initial arrival process has a $v.w.d$ SAC while the departure process has an $i.a.t$ SAC. In order to derive the service guarantees in Server 2, we need Theorem \ref{it2vsdrelation} (2) to transform the $i.a.t$ SAC into a $v.w.d$ SAC. Such transformation introduces a loose bounding function. The node-by-node analysis thus generates a loose end-to-end delay bound. Network calculus possesses an attractive property, \emph{concatenation property}, which is used to deal with the end-to-end performance analysis. The comparison between the node-by-node analysis and the concatenation analysis reveals that the latter yields a tighter end-to-end delay bound \cite{Jiang:TTM4155Notes}.  

\begin{figure}[!thbp]
\centering
\includegraphics[width=0.8\textwidth]{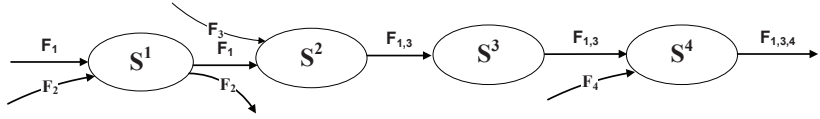}
\caption{Complicated network scenario}
\label{fig:ComplicatedSce}
\end{figure}

The output characterization property however is very useful when analyzing complicated network scenarios, such as Figure \ref{fig:ComplicatedSce}, where flows join or leave dynamically. In order to analyze the per-flow service guarantees, the departure process from each single node should be characterized using the arrival process to the node and the service process provided by the node. 

Moreover, if the arrival process and the service process are independent of each other, the following lemma depicts the departure process. 

\begin{lemma}\label{OutputcharIndependent}
\textbf{(Output Characterization: independent condition.)} 

Consider that a system provides an i.d SSC $\gamma(n)\in\mathcal{F}$ with bounding function $j(x)\in\bar{\mathcal{F}}$ to its arrival process which has a $v.w.d$ SAC $\lambda(n)\in\mathcal{F}$ with bounding function $h(x)\in\bar{\mathcal{F}}$. The output has an i.a.t SAC $\lambda^*\in\mathcal{F}$ with bounding function $h^*(x)\in\bar{\mathcal{F}}$, where
\begin{eqnarray}
\lambda^*(n)=\lambda\bar{\oslash}\gamma(n-1) ~~\text{and}~~
h^*(x)= 1 - \bar{j}*\bar{h}(x).
\label{eq:outputIndependent}
\end{eqnarray}
\end{lemma}

\subsection{Concatenation Property}\label{Concatenation}
The concatenation property aims to use an equivalent system to represent a system of multiple servers connected in tandem if each server provides a service curve to its input. Then this equivalent system can be considered as a \lq black box\rq~which also provides the initial input with a service curve. 

In the following discussion, $\gamma^k$ and $j^k$ denote the stochastic service curve and bounding function of the $k$th server. For packet $P(n)$, the time arriving to the $k$th server is $a^k(n)$ and the time departing from the $k$th server is $d^k(n)$. For a network of $N$ tandem servers, the initial arrival is $a(n)$ and the final departure is $d(n)$.   
\begin{theorem}\label{concatenation}
\textbf{(Concatenation Property)} 

Consider a flow passing through a system of $N$ nodes connected in tandem. If each node $k$(= 1,2,...,N) provides an $i.d$ SSC $\gamma^k(n)\in\mathcal{F}$ with bounding function $j^k(x)\in\bar{\mathcal{G}}$ to its input, the system provides to the initial input $a(n)$ an $i.d$ SSC $\gamma(n)$ with bounding function $j(x)$, where  
\begin{eqnarray}
\gamma(n) &=& \gamma^1\bar{\otimes}\gamma^2_{\eta}\bar{\otimes}\cdot\cdot\cdot\bar{\otimes}\gamma^N_{(N-1)\eta}(n)\nonumber\\
j(x) &=& j^{1,\eta_1}\otimes j^{2,\eta_2}\otimes\cdot\cdot\cdot\otimes j^{N}(x),\nonumber
\end{eqnarray}
with $$\gamma^k_{(k-1)\eta}(n)=\gamma^k(n)+(k-1)\cdot\eta\cdot n$$ for $k=2,...,N$ and $\eta>0$, and $$j^{k,\eta_k}(x)=\big[j^k(x)+\frac{1}{\eta_k}\int_{x}^{\infty}j^k(y)dy\big]_1$$ for $k=1,...,N-1$ and $\eta_k>0$. 
\end{theorem}

The proof of Theorem \ref{concatenation} utilizes the relationship between the $i.d$ SSC and the $\eta-$stochastic service curve. 
The following lemma directly describes the service characterization of a system of nodes connected in tandem, where each single node provides an $\eta-$stochastic service curve to its input.  
\begin{lemma}\label{lemma:concatenation}
Consider a flow passing through a system of $N$ nodes connected in tandem. If each node $k$(= 1,2,...,N) provides an $\eta$-stochastic service curve $\gamma^k(n)\in\mathcal{F}$ with bounding function $j^k(x)\in\bar{\mathcal{F}}$ to its input, i.e., 
\begin{displaymath}
P\Big\{\sup_{0\leq m \leq n}\big\{d^k(m)-a^k\bar{\otimes}\gamma^k(m)-\eta\cdot(n-m)\big\}>x\Big\}\leq j^k(x),
\end{displaymath}
then the system provides to the initial arrival process an $i.d$ SSC $\gamma(n)$ with bounding function $j(x)$:
\begin{displaymath}
\gamma(n) = \gamma^1\bar{\otimes}\gamma^2_{\eta}\bar{\otimes}\cdot\cdot\cdot\bar{\otimes}\gamma^N_{(N-1)\eta}(n)
\end{displaymath} 
\begin{displaymath}
j(x) = j^{1}\otimes j^{2}\otimes\cdot\cdot\cdot\otimes j^{N}(x),
\end{displaymath}
where $\gamma^k_{(k-1)\eta}(n)=\gamma^k(n)+(k-1)\cdot\eta\cdot n$, $k=2,...,N$, for any small $\eta>0$.
\end{lemma}
\textbf{Remark.} The proof of the concatenation property reveals another reason of defining the $\eta-$stochastic service curve model. 

\subsection{Superposition Property}\label{Superposition}
The superposition property can be applied for multiplexing individual flows into an aggregated flow under the FIFO aggregate scheduling. The arrival process of the aggregate flow can be characterized by a stochastic arrival curve if the arrival process of each individual flow can be stochastically characterized by a stochastic arrival curve. Then we only need to analyze the service guarantees for the aggregate flow since all constituent flows are served equally. 

\subsubsection{Superposition of Renewal Processes}\label{Sec-Superposition1}
The superposition of multiple flows essentially falls into the research issue - superposition of renewal processes. In queueing networks, an individual server may receive inputs from different sources. It is reasonable to assume that the arrival process to a server is a superposition of statistically independent constituent processes \cite{Lam:Superposition}. The individual constituent processes are typically considered as renewal processes. A \emph{renewal process} is a counting process in which the times between successive events are independent and identically distributed possibly with an arbitrary distribution \cite{Ross:Probability}. 

The superposition of renewal processes has been widely studied since the original investigation by Cox and Smith \cite{Cox:Superposition}. However, the renewal property is not preserved under superposition except for Poisson sources. More precisely, the inter-arrival times in the superposition process become statistically dependent. This property cannot be captured by the renewal model \cite{Torab:Renewal}. 

In the following, we introduce how to characterize the superposition processes of multiple flows from a network calculus viewpoint.

\subsubsection{Arrival Time Determination}
\begin{figure}[!htbp]
\centering
\includegraphics[width=0.8\textwidth]{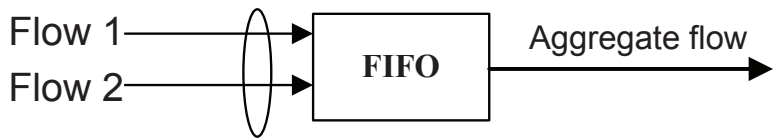}
\caption{Aggregation of two flows}
\label{fig:2flowAgg}
\end{figure}

First, we only consider the superposition of two flows denoted by $F_1$ and $F_2$. Let $a_1(n)$, $a_2(n)$ and $a(n)$ be the arrival process of $F_1$, $F_2$ and the aggregate flow, respectively. As shown in Figure \ref{fig:2flowAgg}, $F_1$ and $F_2$ are aggregated in the FIFO manner. If two or more than two packets which belong to different flows arrive simultaneously, they are inserted into the FIFO queue arbitrarily.    

\begin{figure}[!htbp]
\centering
\includegraphics[width=0.8\textwidth]{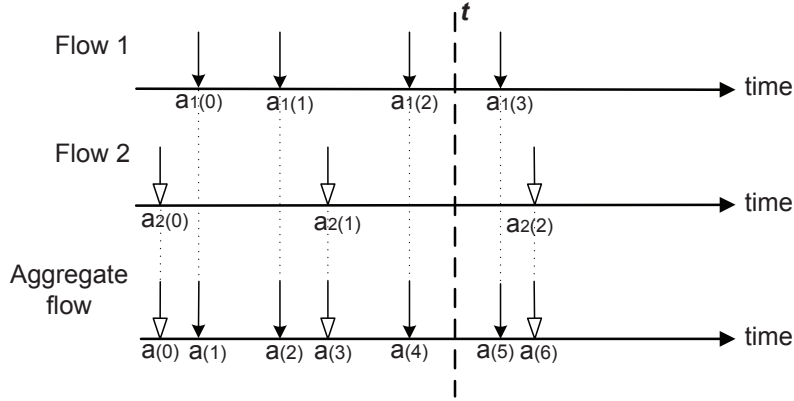}
\caption{Packet arrival time}
\label{fig:Aggpktorder}
\end{figure}

Figure \ref{fig:Aggpktorder} depicts that the arrival process of the aggregate flow is dependent on the arrival process of two constituent flows. 

Recall that $P(n)$ denotes the $(n+1)$th packet of the aggregate flow. The same notation is also used for constituent flows $F_1$ and $F_2$. Thus, 
packet $P(n)$ of the aggregate flow is either the $m$th packet of flow $F_1$ (i.e., $P_1(m-1)$) or the $(n+1-m)$th packet of flow $F_2$ (i.e., $P_2(n-m)$), where $0\le m\leq n+1$. When $m=0$, it means no packet of flow $F_1$ arriving yet. When $m=n+1$, it means no packet of flow $F_2$ arriving yet. By convention, we adopt $a_i(n)=0$ for $n<0$. Since $m$ takes value between $0$ to $n+1$, there are $n+2$ combinations.

\begin{theorem}\label{AggregateArrivalTime}
Consider that two flows $F_1$ and $F_2$ arrive to a network system and are aggregated into one flow $F_A$ in the FIFO manner. Let $a_1(n)$, $a_2(n)$ and $a(n)$ be the arrival process of flows $F_1$, $F_2$ and $F_A$, respectively. Then the packet arrival time of the aggregate flow is determined by  
\begin{equation}
a(n) = \min_{0\leq m \leq n+1}\Big\{\max\big[a_1(m-1),a_2(n-m)\big]\Big\}
\label{eq:finalaggregatearrivaltime}
\end{equation} 
with $$a(0)=\min\Big\{\max\big[0,a_2(0)\big],\max\big[a_1(0),0\big]\Big\}=\min[a_1(0),a_2(0)].$$ 
\end{theorem}

We use an example to explain the underling concept of Theorem \ref{AggregateArrivalTime}. In Figure \ref{fig:Aggpktorder}, observe the arrival process of the aggregate flow at time $t$. Packet $P(4)$ (arrival time: $a(4)<t$) is the last arrival packet, which is either packet $P_1(m-1)$ or packet $P_2(4-m)$, depending on which packet's arrival time is closer to time $t$, i.e., $a(4)=\max[a_1(m-1),a_2(4-m)]$ for $m=0,1,2,3,4,5$, the arrival time of packet $P(4)$ is one element of the following set denoted by $\mathbb{A}$
\begin{eqnarray}
\mathbb{A}&=&\Big\{a_1(4), a_2(4),\max[a_1(0),a_2(3)],\max[a_1(1),a_2(2)], \max[a_1(2),a_2(1)], \nonumber\\
&&\max[a_1(3),a_2(0)]\Big\},\nonumber
\end{eqnarray}
i.e., $a(4)=\min\{\mathbb{A}\}$. We notice that $\min\{\mathbb{A}\}$ is actually the expansion of Eq.(\ref{eq:finalaggregatearrivaltime}). According to the packet arrival times of two constituent flows shown in Figure \ref{fig:Aggpktorder}, we have
\begin{eqnarray}
a(4)&=&\min\Big\{a_1(4), a_2(4),\max[a_1(0),a_2(3)]=a_2(3), \max[a_1(1),a_2(2)]=a_2(2),\nonumber\\
&&\max[a_1(2),a_2(1)]=a_1(2), \max[a_1(3),a_2(0)]=a_1(3)\Big\}=a_1(2),\nonumber
\end{eqnarray}
which is consistent with Figure \ref{fig:Aggpktorder}. 

Theorem \ref{AggregateArrivalTime} can be generalized to the aggregation of $N(\ge 2)$ flows.
\begin{corollary}\label{NflowAggregation}
Consider that $N(\ge 2)$ flows $F_1$,$F_2$,...,$F_N$ arrive to a network system and are aggregated into one flow $F_A$ in the FIFO manner. Let $a_1(n)$, $a_2(n)$,...,$a_N(n)$ and $a(n)$ be the arrival process of the $N$ constituent flows and the aggregate flow, respectively. Then the packet arrival time of the aggregate flow is determined by 
\begin{equation}
a(n) = \min_{\sum m_i=n+1,m_i\in[0,n+1]}\Big\{\max[a_1(m_1-1),a_2(m_2-1),...,a_N(n-\sum_{i=1}^{N-1}m_i)]\Big\}
\label{eq:Naggregate}
\end{equation}
with $$a(0)=\min\big\{a_1(0),a_2(0),...,a_N(0)\big\}.$$
\end{corollary}

\subsubsection{Superposition Process Characterization}
Eq.(\ref{eq:finalaggregatearrivaltime}) can compute the packet arrival time of the aggregate flow. However, we still have the difficulty in characterizing the packet inter-arrival time of the aggregate flow if the packet inter-arrival times of two constituent flows follow the general distribution. 
For this reason, it is difficult to directly characterize the arrival process of the aggregate flow from the temporal perspective. Alternatively, we rely on the available results of the superposition property explored in the space-domain (see Theorem \ref{ThSpaceSuperposition}).  

In the space-domain, the traffic arrival process is characterized based on the cumulative amount of arrival traffic. In the following, we use $\mathcal{A}(t)$, $\mathcal{A}_1(t)$ and $\mathcal{A}_2(t)$ to denote the cumulative number of arrival packets of the aggregate flow up to time $t$, the cumulative number of arrival packets of $F_1$ up to time $t$ and the cumulative number of arrival packets of $F_2$ up to time $t$, respectively. $\mathcal{A}(t)$ is the sum of $\mathcal{A}_1(t)$ and $\mathcal{A}_2(t)$, from which we can find the stochastic arrival curve for the aggregate flow.     
\begin{figure}[!thbp]
\centering
\includegraphics[width=0.8\textwidth]{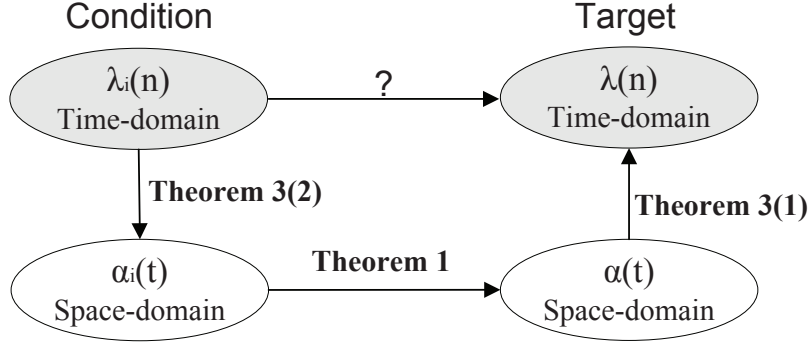}
\caption{Transformation in Theorem \ref{superpositionmaxplus}}
\label{fig:transforsuper}
\end{figure} 

As shown in Figure \ref{fig:transforsuper}, the condition is that the time-domain stochastic arrival curve of all constituent flows are known, and the target is to verify that the aggregate flow also has a time-domain stochastic arrival curve. 

If a flow has a time-domain $v.w.d$ SAC, with Theorem \ref{vbctovsd}(2), this flow has a space-domain $v.b.c$ SAC, for which the superposition property holds (refer Theorem \ref{ThSpaceSuperposition}). Applying Theorem \ref{vbctovsd}(1) gives rise to the $v.w.d$ SAC for the aggregate flow. 

If flow $F_i$ has a $v.w.d$ SAC $\lambda_i(n)$ with bounding function $h_i(x)$, $i=1,2,...,N$, from Theorem \ref{vbctovsd}(2), we can verify that flow $F_i$ has a $v.b.c$ SAC $\alpha_i(t)$ with bounding function $f_i(x)=h_i\big(z^{-1}_i(x)\big)$, where $\alpha_i(t)$ and $z^{-1}_i(x)$ are given in Theorem \ref{vbctovsd}(2). Furthermore, according to Theorem \ref{ThSpaceSuperposition}, the aggregate flow has a $v.b.c$ SAC $\alpha(t)=\sum_{i=1}^N\alpha_i(t)$ with bounding function $f(x)=f_1\otimes\cdot\cdot\cdot\otimes f_N(x)$. Finally, we apply Theorem \ref{vbctovsd}(1) and can verify that the aggregate flow also has a $v.w.d$ SAC.  
\begin{theorem}\label{superpositionmaxplus}
\textbf{(Superposition property)}
 
Consider the aggregate of $N$ flows. If the arrival process of each flow has a $v.w.d$ SAC $\lambda_i(n)\in\mathcal{F}$ for $i=1,2,...,N$, i.e.,
\begin{displaymath}
P\{a_i(n) < a_i\bar{\otimes}\lambda_i(n)-y\}\leq h_i(y),
\end{displaymath}
which implies that every flow also has a $v.b.c$ SAC $$\alpha_i(t)=\sup\{k:\lambda_i(k)\leq t\}$$ with bounding function $$f_i(x)=h_i\big(z^{-1}_i(x)\big)$$ where $z^{-1}_i(x)$ denote the inverse function of $x$: $$x= z_i(y) \equiv \sup_{\tau \ge 0} \{ \alpha_i(\tau+y) - \alpha_i(\tau) +1\}.$$ Then the aggregate arrival process $a(n)$ has a $v.w.d$ SAC $\lambda(n)$ with bounding function $h(y)$, where 
\begin{displaymath}
\lambda(n)=\inf\{\tau:\sum_{i=1}^N\alpha_i(\tau)\geq n\},~~~~h(y)=f\big(z^{-1}(y)\big),  
\end{displaymath}
with $f(x)=f_1\otimes\cdot\cdot\cdot\otimes f_N(x)$ and $z^{-1}(y)$ denoting the inverse function of $y$: $$y= z(x) \equiv \sup_{k\ge 0} \{\lambda(k)-\lambda(k-x)\}.$$ 
\end{theorem}

\subsubsection{Special Case: Superposition of Poisson Processes}\label{Sec-PropertySuperpositionPoisson}
As we have mentioned in Section \ref{Sec-Superposition1}, the Poisson process is a special case of renewal processes because its renewal property is preserved under superposition. In addition, the superposition of multiple Poisson processes is still a Poisson process. From the temporal perspective, the inter-arrival time between two arbitrary events of a superposition of Poisson arrivals follows the Gamma distribution. 

\begin{example}\label{EgPoissonSuperposition}
Consider the superposition process of two independent Poisson arrival processes. Suppose that all packets of both arrival processes have the same size. The packet inter-arrival times of two Poisson processes follow exponential distributions with mean $\frac{1}{\mu_1}$ and $\frac{1}{\mu_2}$, respectively. Find the time-domain $v.w.d$ SAC for the superposition process.    
\end{example}
In Example \ref{examWait} the $v.w.d$ stochastic arrival curve for a Gamma process has been derived. We thus know that the superposition process has a $v.w.d$ SAC $\lambda_s(n)=T_s\cdot n$ ($0<T_s<\frac{1}{\mu_1+\mu_2}$) with bounding function $h_s(x)$: 
\begin{displaymath}
h_s(x) = 1 - (1-\rho_s)\sum_{i=0}^{\lfloor\frac{x}{T_s}\rfloor}e^{-(\mu_1+\mu_2)(iT_s-x)}\frac{[(\mu_1+\mu_2)(iT_s-x)]^i}{i!},
\end{displaymath} 
where $\rho_s=(\mu_1+\mu_2)\cdot T_s$. 

\textbf{Remark.} It is readily to generalize the above example into the superposition of multiple independent Poisson processes.  

\section{Conclusions and Open Issue}\label{Sec-Conclusion}

This paper presented a temporal network calculus to formulate queueing systems in communication networks where applications can tolerate a certain level of performance violation. The time-domain models make it feasible to characterize the temporal behavior of network traffic and capture the temporal nature of the network capacity perceived by individual packets. 

The models are defined in such a way to compromise between simple models and complex models. The former may not be sufficient to explore the fundamental properties whereas the latter may be too difficult to build. In order to solve this dilemma, we propose a transformation method such that the appropriate models are selected to some specific scenario. Moreover, we also link the temporal network calculus and the existing space-domain network calculus results through connecting the time-domain $v.w.d$ arrival curve and the corresponding space-domain $v.b.c$ arrival curve.

Four properties investigated in the time-domain facilitate performance analysis of various network scenarios. In addition, the proof of the superposition property has given insights into the importance of model transformation. We believe that this temporal network calculus is applicable for analyzing networks where users are served probabilistically and compliments the current network calculus results.

The leftover service characterization is useful for per-flow performance analysis and has been proved in the space-domain. We attempted to tackle this property under the condition that the arrival process has a deterministic time-domain arrival curve and the service process provides an $i.d$  SSC \cite{Xie:timedomain}. We will expand the investigation to this property under the general condition. One challenge is to be able to decouple the constituent flow's arrival process from the aggregate arrival process.      
\bibliographystyle{abbrv}
\bibliography{Mybib}

\appendix

\section{Proofs of theorems and lemmas}

\textbf{Proof of Theorem \ref{it2vsdrelation}}
\begin{proof}
The first part follows from that for any $0\leq m \leq n$, there trivially holds
\begin{displaymath}
\lambda(n-m)-[a(n)-a(m)]\leq\sup_{0\leq m\leq n}\big\{\lambda(n-m)-[a(n)-a(m)]\big\}.
\end{displaymath}
For the second part, there holds
\begin{eqnarray}
&&P\Big\{\sup_{0\leq m\leq n}\big\{\lambda_{-\eta}(n-m)-[a(n)-a(m)]\big\}>x\Big\}\nonumber\\
&\leq& P\Big\{ \sup_{0\leq m\leq n}\big\{\lambda_{-\eta}(n-m)-[a(n)-a(m)]\big\}^+>x\Big\}. \nonumber
\end{eqnarray}
For any $x\geq 0$, 
\begin{eqnarray}
&&P\Big\{\{\lambda(n-m)-\eta\cdot(n-m)-[a(n)-a(m)]\}^+>x\Big\}\nonumber\\
&=& P\Big\{\lambda(n-m)-\eta\cdot(n-m)-[a(n)-a(m)] >x\Big\}\nonumber\\
&=& P\Big\{\lambda(n-m)-[a(n)-a(m)] >x+\eta\cdot(n-m)\Big\}\nonumber\\
&\leq& h\big(x+\eta\cdot(n-m)\big).\nonumber
\end{eqnarray}
Based on the above steps, we have
\begin{eqnarray}
&&P\Big\{\sup_{0\leq m\leq n}\{\lambda_{-\eta}(n-m)-[a(n)-a(m)]\}> x\Big\}\nonumber\\
&\leq& \sum_{m=0}^nP\Big\{\{\lambda_{-\eta}(n-m)-[a(n)-a(m)]\}^+> x\Big\}\nonumber\\
&\leq& \sum_{m=0}^nh(x+\eta\cdot(n-m))\nonumber\\
&=&\sum_{k=0}^nh(x+\eta\cdot k)\nonumber\\
&\leq& \sum_{k=0}^{\infty}h(x+\eta\cdot k)\nonumber
\end{eqnarray}
\begin{eqnarray}
&=&h(x)+\sum_{k=1}^{\infty}h(x+\eta\cdot k)\nonumber\\
&\leq& h(x)+\frac{1}{\eta}\int_{x}^{\infty}h(y)dy.\nonumber
\end{eqnarray}
The right-hand side of the last inequality still belongs to $\bar{\mathcal{G}}$. The second part follows from the above inequality and the fact that the probability is always not greater than $1$. 
\end{proof}

\textbf{Proof of Theorem \ref{vbctovsd}}
\begin{proof}
(1) From Lemma 2 \cite{Xie:timedomain}, we know that for any $t, x\geq 0$, event $$\{\mathcal{A}(t)\leq\mathcal{A}\otimes\alpha(t)+x\}$$ implies event $$\{a(n)\geq a\bar{\otimes}\lambda(n)-y\}$$ where $$y= \sup_{k\ge 0} \big\{\lambda(k)-\lambda(k-x)\big\} \equiv z(x).$$
 Thus, there holds 
\begin{eqnarray}
P\big\{\mathcal{A}(t)\leq\mathcal{A}\otimes\alpha(t)+x\big\}&\leq& P\big\{a(n)\geq a\bar{\otimes}\lambda(n)-y\big\}\nonumber\\
\Longrightarrow ~~~~P\big\{a(n)<a\bar{\otimes}\lambda(n)-y\big\} &\leq& P\big\{\mathcal{A}(t)>\mathcal{A}\otimes\alpha(t)+x\big\}.\nonumber\\
&\leq& f(x),\nonumber
\end{eqnarray}
Particularly, if $\lambda$ is sub-additive, i.e. $\lambda(a+b) \le \lambda(a) + \lambda(b)$ for any $a$ and $b$, we then have:
\begin{eqnarray}
&&P\big\{a(n)<a\bar{\otimes}\lambda(n)-\lambda(x)\big\} \nonumber\\
&\leq& P\big\{a(n)<a\bar{\otimes}\lambda(n)-\sup_{k\geq 0}[\lambda(k)-\lambda(k-x)]\big\}\nonumber\\
&\leq& f(x).\nonumber
\end{eqnarray}
Hence, the first part follows.

(2) From Lemma 3 \cite{Xie:timedomain}, we know that for any $n, y\geq 0$, event $$\{a(n)\geq a\bar{\otimes}\lambda(n)-y\}$$ implies event $$\{\mathcal{A}(t)\leq\mathcal{A}\otimes\alpha(t)+x\}$$ where $$x=\sup_{u \ge 0} \{ \alpha(u+y) - \alpha(u) +1\} \equiv z(y).$$ 
Thus, there holds
\begin{eqnarray}
P\big\{a(n)\geq a\bar{\otimes}\lambda(n)-y\big\}&\leq& P\big\{\mathcal{A}(t)\leq\mathcal{A}\otimes\alpha(t)+x\big\}\nonumber\\
\Longrightarrow ~~~~P\big\{\mathcal{A}(t)>\mathcal{A}\otimes\alpha(t)+x\big\} &\leq& P\big\{a(n)< a\bar{\otimes}\lambda(n)-y\big\}\nonumber\\
&\leq& h(y).\nonumber
\end{eqnarray}
Particularly, if $\alpha$ is sub-additive, we have 
\begin{eqnarray}
&&P\big\{\mathcal{A}(t)>\mathcal{A}\otimes\alpha(t)+ \alpha(y) +1\big\} \nonumber\\
&\le& P\big\{\mathcal{A}(t)>\mathcal{A}\otimes\alpha(t)+ \sup_{u \ge 0}[\alpha(u+y) - \alpha(u) +1]\big\} \nonumber\\
&\le& h(y),\nonumber
\end{eqnarray}
which ends the proof.
\end{proof}

\textbf{Proof of Theorem \ref{stosercur2constosercur}}


\begin{proof}
The first part follows since there always holds 
\begin{displaymath}
d(n)-a\bar{\otimes}\gamma(n)\leq \sup_{0\leq m\leq n}\big[d(m)-a\bar{\otimes}\gamma(m)-\eta\cdot(n-m)\big]
\end{displaymath}
by letting $m=n$ on the right hand side. 

For the second part, there holds 
\begin{eqnarray}
&&\sup_{0\leq m \leq n}\big[d(m) - a\bar{\otimes}\gamma(m)-\eta\cdot(n-m)\big] \nonumber\\
&\leq& \sup_{0\leq m \leq n}\{d(m) - a\bar{\otimes}\gamma(m)-\eta\cdot(n-m)\}^+. \nonumber
\end{eqnarray}
Hence for any $x\geq 0$, there exists

\begin{eqnarray}
&&P\big\{\sup_{0\leq m\leq n}\{d(m)-a\bar{\otimes}\gamma(m)-\eta\cdot(n-m)\} > x\big\}\nonumber\\
&\leq& \sum_{m=0}^n P\big\{ d(m)-a\bar{\otimes}\gamma(m) - \eta\cdot(n-m) > x \big\}\nonumber
\end{eqnarray}
\begin{eqnarray}
&\leq& \sum_{u=0}^{n} j(x+\eta\cdot u)\nonumber\\
&\leq& \Big[j(x) +\frac{1}{\eta}\int_{x}^{\infty} j(y)dy\Big]_1.\nonumber
\end{eqnarray}
The right-hand side of the above inequality still belongs to $\bar{\mathcal{G}}$ and is always not greater than 1. The proof of the second part is completed. 
\end{proof}

\textbf{Proof of Theorem \ref{Delaybound}}
\begin{proof}
For any $n\geq 0$, according to the definition of $D(n)$, there holds 
\begin{eqnarray}
D(n)&=& d(n) - a(n)\nonumber\\
 &=& \big[d(n) - a\bar{\otimes}\gamma(n)\big] + \big[a\bar{\otimes}\gamma(n) - a(n)\big]\nonumber\\
&=& \big[d(n) - a\bar{\otimes}\gamma(n)\big] + \sup_{0\le m\le n}\big\{a(m)+\gamma(n-m+1)-a(n)\big\}\nonumber\\
&=& \big[d(n) - a\bar{\otimes}\gamma(n)\big] + \sup_{0 \leq m \leq n}\big\{\lambda(n-m) - [a(n) - a(m)] + \nonumber\\
&&~~~~~~~~~~~~~~~~~~~~~~~~~~~~~~~~~~~~\gamma(n-m+1) - \lambda(n-m)\big\}\nonumber\\
&\leq& d(n) - a\bar{\otimes}\gamma(n)+\sup_{0\leq m\leq n}\big\{\lambda(n - m) - [a(n) - a(m)]\big\}\nonumber\\
&&+\sup_{0\leq m\leq n}\{\gamma(n-m+1)-\lambda(n-m)\}\nonumber\\
&\leq& d(n) - a\bar{\otimes}\gamma(n)+\sup_{0\leq m\leq n}\big\{\lambda(n - m) - [a(n) - a(m)]\big\} \nonumber\\
&&+ \sup_{k\geq 0}\{\gamma(k+1) - \lambda(k)\}.\nonumber
\end{eqnarray}
To ensure system stability, we require
\begin{equation}
~~~~~~~~~~\lim_{k\to \infty}\frac{1}{k}[\gamma(k) - \lambda(k)] \leq 0.
\label{eq:systemstable}
\end{equation} 
In the proofs of the following theorems, without explicitly stating, we shall assume Eq.(\ref{eq:systemstable}) holds. 

In addition, the following results are given $$P\big\{d(n) - a\bar{\otimes}\gamma(n) > x\big\} ~~\text{and}~~ P\Big\{\sup_{0\leq m\leq n}\big\{\lambda(n - m) - \big[a(n) - a(m)\big]\big\} > x\Big\}.$$ From Lemma 1.5 \cite{Jiang:book} and $\sup_{k\geq 0}\big\{\gamma(k+1)-\lambda(k)\big\}=\gamma\oslash\lambda(1),$ we can conclude 
\begin{displaymath}
P\{D(n) > x\} \leq j\otimes h(x-\gamma\oslash\lambda(1)).~~~~~~~~~~~~~~~~
\end{displaymath}
\end{proof}

\textbf{Proof of Theorem \ref{BacklogBound}}
\begin{proof}
According to the backlog definition $$B(t)\le \inf\Big\{x\ge 0, \sup\{n\ge 0: a(n)\le t\}: d(n-x)\le a(n)\Big\},$$ we need to prove the bounding function on the violation probability, i.e., $P\{B(t)>x\}$. For ease of exposition, let $n=m+x$, then we have 
\begin{eqnarray}
&&d(m) - a(m+x) \nonumber\\
&=& \big[d(m) - a\bar{\otimes}\gamma(m)\big] + \big[a\bar{\otimes}\gamma(m) - a(m+x)\big]\nonumber\\
&=& \big[d(m) - a\bar{\otimes}\gamma(m)\big] + \sup_{0\le k\le m}\big\{a(k)+\gamma(m-k+1)\big\}-a(m+x)\nonumber\\
&=& \big[d(m) - a\bar{\otimes}\gamma(m)\big] +\sup_{0\leq k\leq m}\big\{\lambda(m+x-k) - [a(m+x) - a(k)]\nonumber\\
&&  ~~~~~~~~~~~~~~~~~~~~~~~~~~~~~~~~~~~~~~+ \gamma(m-k+1)-\lambda(m+x-k)\big\} \nonumber\\
&\leq& \big[d(m) - a\bar{\otimes}\gamma(m)\big] + \sup_{0\leq k\leq m+x}\big\{\lambda(m+x-k) - [a(m+x)- a(k)]\big\}\nonumber\\
 && - \inf_{0\leq k\leq m}\big\{\lambda(m-k+x) - \gamma(m-k+1)\big\}\nonumber
\end{eqnarray}
Let $v = m-k+1$. The above inequality is written as
\begin{eqnarray}
&&d(m) - a(m+x)\nonumber\\
& \leq& \big[d(m) - a\bar{\otimes}\gamma(m)\big] + \sup_{0\leq k\leq m+x}\big\{\lambda(m+x-k) - [a(m+x) - a(k)]\big\}\nonumber\\
&& - \inf_{1\leq v\leq m+1}\big\{\lambda(v+x-1) - \gamma(v)\big\}.\nonumber
\end{eqnarray}
Because there holds
\begin{eqnarray}
\inf_{1\leq v\leq m+1}\big\{\lambda(v+x-1) - \gamma(v)\big\}&\geq& \inf_{v\geq 1}\big\{\lambda(v+x-1) - \gamma(v)\big\}\nonumber\\
&=&\lambda\bar{\oslash}\gamma([x-1]^+),\nonumber
\end{eqnarray}
with the same conditions as analyzing the delay, we obtain
\begin{displaymath}
P\{B(t) > x\} \leq j\otimes h\big(\lambda\bar{\oslash}\gamma([x-1]^+)\big).
\end{displaymath}
To prove Eq.(\ref{eq:maxBtbound}), we replace $x=H(\lambda,\gamma+y)+1$ in event $\{B(t) > x\}$ and have
\begin{eqnarray}
&&d(m)-a(m+H(\lambda,\gamma+y)+1) \nonumber\\
&\leq& \big[d(m) - a\bar{\otimes}\gamma(m)\big] + a\bar{\otimes}\lambda\big(m+H(\lambda,\gamma+y)+1\big) \nonumber\\
&&-a\big(m+H(\lambda,\gamma+y)+1\big)+ \sup_{v\geq 0}\big\{\gamma(v)-\lambda(v+H(\lambda,\gamma+y))\big\}.\nonumber
\end{eqnarray}
The definition of $H(\lambda,\gamma+y)$ implies $$\gamma(v)+y \leq\lambda(v+H(\lambda,\gamma+y))$$ for any $v\geq 0$, i.e., $$\sup_{v\geq 0}\big\{\gamma(v)-\lambda(v+H(\lambda,\gamma+y))\big\} \leq -y.$$ Then we conclude 
\begin{displaymath}
P\{B(t) > H(\lambda,\gamma+x)+1\} \leq j\otimes h(x).
\end{displaymath} 
\end{proof}

\textbf{Proof of Theorem \ref{Outputchar1}}

\begin{proof}
For any two departure packets $m<n$, there holds 
\begin{eqnarray}
d(m) - d(n) &\leq& d(m) - a(n)\nonumber\\ 
&=& d(m) - a(n) + a\bar{\otimes}\gamma(m) - a\bar{\otimes}\gamma(m)\nonumber\\
&=& \big[d(m) - a\bar{\otimes}\gamma(m)\big]+\sup_{0\le k\le m}\big\{a(k)+\gamma(m-k+1)\big\}-a(n)\nonumber\\
&=& \big[d(m) - a\bar{\otimes}\gamma(m)\big]+\sup_{0\le k\le m}\big\{\gamma(m-k+1)-[a(n)-a(k)]\big\}\nonumber\\
&=&\big[d(m) - a\bar{\otimes}\gamma(m)\big]+\sup_{0\le k\le m}\Big\{\gamma(m-k+1)-\lambda(n-k)\nonumber\\
&&~~~~~~~~~~~~~~~~~~~~~~~~~~~~~~~~~~~~+\lambda(n-k)-[a(n)-a(k)]\Big\}\nonumber\\
&\leq& \big[d(m) - a\bar{\otimes}\gamma(m)\big] + \sup_{0\leq k\leq m}\Big\{\lambda(n-k) - [a(n) - a(k)]\Big\}\nonumber\\
&&+ \sup_{0\leq k \leq m}\big\{\gamma(m-k+1) - \lambda(n-k)\big\}.\nonumber
\end{eqnarray}
Let $v=m-k+1$. Then the above inequality is written as
\begin{eqnarray}
d(m)-d(n)&\le&  \big[d(m) - a\bar{\otimes}\gamma(m)\big] + \sup_{0\leq k\leq n}\Big\{\lambda(n-k) - [a(n) - a(k)]\Big\}\nonumber\\
&&- \inf_{1\leq v \leq m+1}\big\{\lambda(n - m-1+ v) - \gamma(v)\big\}\nonumber\\
&\le& \big[d(m) - a\bar{\otimes}\gamma(m)\big] + \sup_{0\leq k\leq n}\Big\{\lambda(n-k) - [a(n) - a(k)]\Big\}\nonumber\\
&&- \inf_{0\leq v \leq m+1}\big\{\lambda(n - m-1+ v) - \gamma(v)\big\} \nonumber 
\end{eqnarray}
where the last step is because $$\inf_{0\le k\le m+1}[f_k]\le\inf_{1\le k\le m+1}[f_k].$$ 
Adding $\inf_{0\leq v \leq m+1}\big\{\lambda(n - m -1+ v) - \gamma(v)\big\}$ to both sides of the above inequality results in
\begin{eqnarray}
&&\inf_{0 \leq v \leq m+1}\big\{\lambda(n - m -1+ v) - \gamma(v)\big\} - [d(n) - d(m)]\nonumber\\
&\leq& \big[d(m) - a\bar{\otimes}\gamma(m)\big] + \sup_{0\leq k\leq n}\big\{\lambda(n-k) - [a(n) - a(k)]\big\}.    \nonumber
\end{eqnarray}
In addition, there holds 
\begin{eqnarray}
\lambda\bar{\oslash}\gamma(n-m-1)&=&\inf_{v\ge0}\big\{\lambda(n-m-1+v)-\gamma(v)\big\}\nonumber\\
&\le&\inf_{0\le v\le m+1}\big\{\lambda(n-m-1+v)-\gamma(v)\big\}.\nonumber
\end{eqnarray}
To ensure that the right-hand side of the above inequality is meaningful, it requires $n-m-1>0$. 
With the same conditions as analyzing delay, we conclude
\begin{eqnarray}
&&P\Big\{\lambda\bar{\oslash}\gamma(n-m-1) - [d(n) - d(m)] > x\Big\} \nonumber\\
&\le& P\Big\{\big[d(m) - a\bar{\otimes}\gamma(m)\big] + \sup_{0\leq k\leq n}\big\{\lambda(n-k) - [a(n) - a(k)]\big\} > x\Big\}\nonumber\\
&\le& j\otimes h(x).\nonumber
\end{eqnarray}
\end{proof}
\newpage
\textbf{Proof of Theorem \ref{concatenation}}
\begin{proof}
We shall only prove the three-node case, from which, the proof can be easily extended to the $N$-node case. The departure of the first node is the arrival to the second node, so $d^1(n) = a^2(n)$ and $d^2(n)=a^3(n)$. We then have, 
\begin{eqnarray}
&&d(n)-a\bar{\otimes}\gamma^1\bar{\otimes}\gamma^2_{\eta}\bar{\otimes}\gamma^3_{2\eta}(n) \nonumber\\
&=&d(n)-\sup_{0\le m\le n}\Big\{a\bar{\otimes}\gamma^1(m)+\gamma^2_{\eta}\bar{\otimes}\gamma^3_{2\eta}(n-m+1)\Big\}+d^1(m)-d^1(m)\nonumber\\
&\le& d(n)-\sup_{0\leq m \leq n}\Big\{\gamma^2_{\eta}\bar{\otimes}\gamma^3_{2\eta}(n-m+1)+d^1(m)-\eta\cdot(n-m+1)\nonumber\\
&&~~~~~~~~~~~~~~~~~~~~~-[d^1(m)-a\bar{\otimes}\gamma^1(m)-\eta\cdot(n-m)]\Big\}\nonumber\\
&\le& d(n)-\sup_{0\leq m \leq n}\Big\{\gamma^2_{\eta}\bar{\otimes}\gamma^3_{2\eta}(n-m+1)+d^1(m)-\eta\cdot(n-m+1)\Big\}\nonumber\\
&&+\sup_{0\le m\le n}\Big\{d^1(m)-a\bar{\otimes}\gamma^1(m)-\eta\cdot(n-m)]\Big\}\nonumber\\
&=& d(n)-\sup_{0\leq m \leq n}\{a^2(m)+\sup_{0\leq k\leq n-m+1}[\gamma^2(k)+\eta\cdot k+\gamma^3(n-m+1-k)~\nonumber\\
&&+2\eta\cdot(n-m+1-k)] -\eta\cdot(n-m+1)\big\}+\nonumber\\
&&+\sup_{0\leq m\leq n}\big\{d^1(m)-a\bar{\otimes}\gamma^1(m)-\eta\cdot(n-m)\big\}\nonumber\\
&=& d(n)-\sup_{0\leq m \leq n}\big\{a^2(m)+\sup_{0\leq k\leq n-m+1}[\gamma^2(k)+\gamma^3(n-m+1-k)\nonumber\\
&&+\eta\cdot(n-m+1-k)]\big\}+ \sup_{0\leq m\leq n}\big\{d^1(m)-a\bar{\otimes}\gamma^1(m)-\eta\cdot(n-m)\big\}\nonumber\\
&=&d(n)- a^2\bar{\otimes}\gamma^2\bar{\otimes}\gamma^3_{\eta}(n) +\sup_{0\leq m\leq n}\big\{d^1(m)-a\bar{\otimes}\gamma^1(m)-\eta\cdot(n-m)\big\}\nonumber\\
&\le& d(n)-\sup_{0\le m\le n}\Big\{a^2\bar{\otimes}\gamma^2(m)+\gamma^3_\eta(n-m+1)\Big\}-a^3(m)+\eta\cdot(n-m+1)\nonumber\\
&& +d^2(m)-\eta\cdot(n-m) + \sup_{0\leq m\leq n}\big\{d^1(m)-a\bar{\otimes}\gamma^1(m)-\eta\cdot(n-m)\big\}\nonumber\\
&\leq& d(n)-\sup_{0\leq m\leq n}\big\{a^3(m)+\gamma^3_{\eta}(n-m+1)-\eta\cdot(n-m+1)\big\}\nonumber\\
&&+\sup_{0\leq m\leq n}\big\{d^2(m)-a^2\bar{\otimes}\gamma^2(m)-\eta\cdot(n-m)\big\}\nonumber\\
&&+ \sup_{0\leq m\leq n}\big\{d^1(m)-a\bar{\otimes}\gamma^1(m)-\eta\cdot(n-m)\big\}\nonumber
\end{eqnarray}
\begin{eqnarray}
&=& d(n) - a^3\bar{\otimes}\gamma^3(n) + \sup_{0\leq m\leq n}\big\{d^2(m)-a^2\bar{\otimes}\gamma^2(m)-\eta\cdot(n-m)\big\}\nonumber\\
&&+ \sup_{0\leq m\leq n}\big\{d^1(m)-a\bar{\otimes}\gamma^1(m)-\eta\cdot(n-m)\big\}.\nonumber
\end{eqnarray}
 
Based on the relationship between the $i.d$ SSC and the $\eta$-stochastic service curve presented in Theorem \ref{stosercur2constosercur}(2), the following inequality holds
\begin{displaymath}
P\{d(n) - a\bar{\otimes}\gamma^1\bar{\otimes}\gamma^2_{\eta}\bar{\otimes}\gamma^3_{2\eta}(n) > x\}\leq j^3\otimes j^{2,\eta_2}\otimes j^{1,\eta_1},
\end{displaymath}
which completes the proof.

Note that both the max-plus convolution and the min-plus convolution are associative and commutative. 
\end{proof}

\textbf{Proof of Lemma \ref{lemma:concatenation}}
\begin{proof}
We shall only prove two-node case, from which, the proof can be extended to the $N$-node case. Keep in mind that $a^2(n)=d^1(n)$. For the two-node case, we have 
\begin{eqnarray}
&&d(n) - a\bar{\otimes}\gamma^1\bar{\otimes}\gamma_{\eta}^2(n)\nonumber\\
&=& d(n)-\sup_{0\leq m \leq n}\big\{a\bar{\otimes}\gamma^1(m)+\gamma^2(n-m+1)+\eta\cdot(n-m+1)\big\}\nonumber\\
&\le& d(n)-\sup_{0\leq m \leq n}\big\{a\bar{\otimes}\gamma^1(m)+\gamma^2(n-m+1)+\eta\cdot(n-m)\big\}\nonumber\\
&&+d^1(m)-a^2(m)\nonumber\\
&\le&d(n)-\sup_{0\leq m \leq n}\big\{a^2(m)+\gamma^2(n-m+1)\big\}\nonumber\\
&&+\sup_{0\le m\le n}\big\{d^1(m)-a\bar{\otimes}\gamma^1(m)-\eta\cdot(n-m)\big\}\nonumber\\
&=& d(n)-a^2\bar{\otimes}\gamma^2(n)+\sup_{0\leq m\leq n}\big\{d^1(m)-a\bar{\otimes}\gamma^1(m)-\eta\cdot(n-m)\big\}\nonumber\\
&\leq& \sup_{0\leq m\leq n}\big\{d(m)-a^2\bar{\otimes}\gamma^2(m)-\eta\cdot(n-m)\big\} \nonumber\\
&&+  \sup_{0\leq m\leq n}\big\{d^1(m)-a\bar{\otimes}\gamma^1(m)-\eta\cdot(n-m)\big\}.\nonumber
\end{eqnarray}
The last step holds because of Theorem \ref{stosercur2constosercur}(1). From the condition, we conclude 
\begin{eqnarray}
P\big\{d(n)- a\bar{\otimes}\gamma^1\bar{\otimes}\gamma_{\eta}^2(n)>x\big\}\le j^1\otimes j^2(x).\nonumber
\end{eqnarray}
\end{proof}

\textbf{Proof of Theorem \ref{AggregateArrivalTime}}
\begin{proof}
We use the induction way to prove this theorem. 

Step (1) We start from $n=1$ with the given condition $a(0)=\min[a_1(0),a_2(0)]$. If $a(0)=a_1(0)$, then $a(1)=\min\big\{a_1(1),a_2(0)\big\}$; if $a(0)=a_2(0)$, then $a(1)=\min\big\{a_1(0),a_2(1)\big\}$.

We expand Eq.(\ref{eq:finalaggregatearrivaltime}) into the following expression:
\begin{eqnarray}
a(1)&=&\min\big\{a_1(1),a_2(1),\max[a_1(0),a_2(0)]\big\}\nonumber\\
&=& \begin{cases}
 \min\big\{a_1(1),a_2(0)\big\}, & \text{if}~~ \min[a_1(0),a_2(0)]=a_1(0);\\
\min\big\{a_1(0),a_2(1)\big\}, & \text{if}~~ \min[a_1(0),a_2(0)]=a_2(0).
\end{cases}\nonumber
\end{eqnarray}
Thus Eq.(\ref{eq:finalaggregatearrivaltime}) holds for $n=1$.

Step (2) Assume $n=k$ holds for $k>1$: 
\begin{displaymath}
a(k) = \min_{0\leq m \leq k+1}\Big\{\max\big[a_1(m-1),a_2(k-m)\big]\Big\} ~~\text{ (induction hypothesis)},
\end{displaymath}
which has four solutions as below:
\begin{eqnarray}
a(k) &=& \begin{cases}
a_1(k), & \text{if} ~~a_1(k) < a_2(0);\\
a_2(k), & \text{if} ~~a_2(k) < a_1(0);\\
a_1(m^*-1), & \text{if} ~~ a_2(k-m^*) < a_1(m^*-1) ~~\text{for}~~ 0< m^* <k+1;\\
a_2(k-m^*), & \text{if} ~~ a_1(m^*-1) < a_2(k-m^*)~~\text{for}~~ 0< m^* <k+1.
\end{cases}\nonumber
\end{eqnarray}

Step (3) Prove $n=k+1$ holds: $$a(k+1) = \min_{0\leq m \leq (k+1)+1}\Big\{\max\big[a_1(m-1),a_2(k+1-m)\big]\Big\}$$  
which can be expanded into 
\begin{eqnarray}
a(k+1)&=&\min\Big\{a_1(k+1), a_2(k+1), \max\big[a_1(0),a_2(k)\big],\max\big[a_1(1),a_2(k-1)\big],\nonumber\\
&& \max\big[a_1(2),a_2(k-2)\big],...,\max\big[a_1(k),a_2(0)\big]\Big\}.\label{eq:proofInduction1}
\end{eqnarray}
We prove Eq.(\ref{eq:proofInduction1}) based on the four solutions of the induction hypothesis, respectively.
\begin{enumerate}[(I)]
  \item If $a(k)=a_1(k)$ which implies $a_1(k) < a_2(0)$, then $a(k+1)$ is either $a_1(k+1)$ or $a_2(0)$ depending on the minimum one, i.e., $a(k+1)=\min\big\{a_1(k+1),a_2(0)\big\}$. Since the condition $a_1(k)<a_2(0)$ implies 
$$a_1(0)<a_1(1)<...<a_1(k)<a_2(0)<a_2(1)...<a_2(k),$$
the induction hypothesis given in Step (2) is expanded into $$a(k)=\min\big\{a_1(k),a_2(k),a_2(k-1),...,a_2(0)\big\}=a_1(k).$$ From this, Eq.(\ref{eq:proofInduction1}) becomes 
\begin{eqnarray}
a(k+1) &=& \min\big\{a_1(k+1),a_2(k+1),a_2(k),a_2(k-1),...,a_2(0)\big\} \nonumber\\
&=& \min\big\{a_1(k+1),a_2(0)\big\},\nonumber
\end{eqnarray}
which proves that Eq.(\ref{eq:proofInduction1}) holds.
  \item If $a(k)=a_2(k)$ which implies $a_2(k)<a_1(0)$, then $a(k+1)$ is either $a_2(k+1)$ or $a_1(0)$ depending on the minimum one, i.e., $a(k+1)=\min\big\{a_2(k+1),a_1(0)\big\}$. Since the condition $a_2(k)<a_1(0)$ implies 
$$a_2(0)<a_2(1)<...<a_2(k)<a_1(0)<a_1(1)...<a_1(k),$$
the induction hypothesis given in Step (2) is expanded into $$a(k)=\min\big\{a_2(k),a_1(k),a_1(k-1),...,a_1(0)\big\}=a_2(k).$$ From this, Eq.(\ref{eq:proofInduction1}) becomes 
\begin{eqnarray}
a(k+1) &=& \min\big\{a_2(k+1),a_1(k+1),a_1(k),a_1(k-1),...,a_1(0)\big\}\nonumber\\
&=& \min\big\{a_2(k+1),a_1(0)\big\},\nonumber
\end{eqnarray}
which proves that Eq.(\ref{eq:proofInduction1}) holds.

\item Without loss of generality, if $a(k) = a_1(m^*-1)$ for $0< m^* < k+1$, which implies $a_2(k-m^*)<a_1(m^*-1)$, then $a(k+1)$ is either $a_1(m^*)$ or $a_2(k-m^*+1)$ depending on the minimum one, i.e., $a(k+1)=\min\big\{a_1(m^*),a_2(k+1-m^*)\big\}$. Since the condition $a_2(k-m^*)<a_1(m^*-1)$ implies $a_2(k-m^*)<a_1(m^*-1)<a_1(m^*)\le a_1(k)$ and $a_2(k-m^*)<a_1(m^*-1)<a_2(k+1-m^*)\le a_2(k)$ for $0<m^*<k+1$, the induction hypothesis given in Step (2) is expanded into 
\begin{eqnarray}
a(k)&=&\min\big\{a_1(k),a_2(k),a_2(k-1),...,a_2(k+1-m^*),a_1(m^*-1),...,\nonumber \\
&&a_1(k-1)\big\} =a_1(m^*-1) \nonumber
\end{eqnarray}
From this, Eq.(\ref{eq:proofInduction1}) becomes 
\begin{eqnarray}
a(k+1) &=& \min\big\{a_1(k+1),a_2(k+1),a_2(k),a_2(k-1),...,\nonumber\\
&&a_2(k+1-m^*),a_1(m^*),a_1(m^*+1),...,a_1(k)\big\}\nonumber\\
&=&\min\big\{a_2(k+1-m^*),a_1(m^*)\big\}\nonumber
\end{eqnarray}
which proves that Eq.(\ref{eq:proofInduction1}) holds. 

\item Without loss of generality, if $a(k) = a_2(k-m^*)$ for $0< m^* < k+1$, which implies $a_1(m^*-1)<a_2(k-m^*)$, then $a(k+1)$ is either $a_1(m^*)$ or $a_2(k-m^*+1)$ depending on the minimum one, i.e., $a(k+1)=\min\big\{a_1(m^*),a_2(k+1-m^*)\big\}$. Since the condition $a_1(m^*-1)<a_2(k-m^*)$ implies $a_1(m^*-1)<a_2(k-m^*)<a_1(m^*)\le a_1(k)$ and $a_1(m^*-1)<a_2(k-m^*)<a_2(k+1-m^*)\le a_2(k)$ for $0<m^*<k+1$, the induction hypothesis given in Step (2) is expanded into 
\begin{eqnarray}
a(k)&=&\min\big\{a_1(k),a_2(k),a_2(k-1),...,a_2(k-m^*),a_1(m^*),...,\nonumber \\
&&a_1(k-1)\big\} =a_2(k-m^*). \nonumber
\end{eqnarray}
From this, Eq.(\ref{eq:proofInduction1}) becomes 
\begin{eqnarray}
a(k+1) &=& \min\big\{a_1(k+1),a_2(k+1),a_2(k),a_2(k-1),...,\nonumber\\
&&a_2(k+1-m^*),a_1(m^*),a_1(m^*+1),...,a_1(k)\big\}\nonumber\\
&=&\min\big\{a_2(k+1-m^*),a_1(m^*)\big\}\nonumber
\end{eqnarray}
which proves that Eq.(\ref{eq:proofInduction1}) holds. 
\end{enumerate}
Combining the above three steps concludes that Eq.(\ref{eq:finalaggregatearrivaltime}) holds for all $n\ge 0$. 
\end{proof}





\end{document}